\documentclass[sigconf]{acmart}
\usepackage{tabularx}
\usepackage{algorithm}
\usepackage{algorithmic}
\usepackage{etoolbox}
\usepackage{xcolor}
\usepackage{threeparttable} 
\newtoggle{showadditions}
\newtoggle{showdeletions}
\toggletrue{showadditions}
\togglefalse{showdeletions}
\newcolumntype{C}[1]{>{\centering\let\newline\\\arraybackslash\hspace{0pt}}m{#1}}

\AtBeginDocument{%
  }

\copyrightyear{2026}
\acmYear{2026}
\setcopyright{cc}
\setcctype{by}
\acmConference[ASSETS '26]{The 28th International ACM SIGACCESS Conference on Computers and Accessibility}{October 25--28, 2026}{Vila Nova de Gaia, Portugal}
\acmBooktitle{The 28th International ACM SIGACCESS Conference on Computers and Accessibility (ASSETS '26), October 25--28, 2026, Vila Nova de Gaia, Portugal}
\acmDOI{10.1145/3797867.3829061}
\acmISBN{979-8-4007-2521-0/2026/10}

\begin{document}

\title[Understanding ADHD Video Learning]{``It's Like Drinking from a Fire Hose'': Understanding and Characterizing Video Learning Experiences for Individuals with ADHD}


\author{Hanxiu `Hazel' Zhu}
\affiliation{%
  \institution{University of Wisconsin-Madison}
  \city{Madison}
  \state{Wisconsin}
  \country{USA}
}
\email{hzhu339@wisc.edu}

\author{Weiyu Zhang}
\affiliation{%
  \institution{University of Wisconsin-Madison}
  \city{Madison}
  \state{Wisconsin}
  \country{USA}
  }
\email{wzhang769@wisc.edu}

\author{Ru Wang}
\affiliation{%
  \institution{University of Wisconsin-Madison}
  \city{Madison}
  \state{Wisconsin}
  \country{USA}
}
\email{ru.wang@wisc.edu}

\author{Yuhang Zhao}
\affiliation{%
  \institution{University of Wisconsin-Madison}
  \city{Madison}
  \state{Wisconsin}
  \country{USA}
  }
\email{yuhang.zhao@cs.wisc.edu}

\renewcommand{\shortauthors}{Zhu et al.}

\begin{abstract}
Video lectures have become increasingly prevalent for education and professional development, yet their static visuals, dense information, and long duration pose attentional challenges for individuals with ADHD. While adaptive learning offers opportunities towards ADHD-accessible video learning, little is known about how to suitably adapt such videos: What components in multimodal video lectures are challenging for ADHD viewers? How do these experiences surface in behavioral signals to trigger an adaptation? What presentations do they prefer? To answer these questions, we conducted an eye-tracking-based retrospective think-aloud study with 16 participants with ADHD, who watched and reflected on a curated set of video lecture segments. Our study uncovered video design elements that hindered learning 
and revealed participants' coping strategies 
along with their limitations. 
By jointly analyzing behavioral signals and retrospective reflections, we characterized how these experiences manifested in behavioral patterns. 
We further surfaced participants' practices for addressing learning needs beyond the video watching process, and derived design implications for future ADHD-friendly adaptive video learning systems.
\end{abstract}

%
%





\maketitle

\section{Introduction}
Video lectures present both opportunities and challenges for individuals with Attention Deficit Hyperactivity Disorder (ADHD) \cite{doernberg2016neurodevelopmental, reaser2007learning}. On the one hand, video lectures afford convenience, flexibility, and sometimes built-in video playback features (e.g., pause/seek, captions, playback speed adjustment), allowing individuals with ADHD to engage with learning materials at their preferred locations, times, and pace \cite{levenberg2023learning}. On the other hand, however, video lectures---with their relatively static visuals, long durations, and dense information structures \cite{emam2025enhancing, alpert2019video}---can be uniquely challenging for individuals with ADHD, who often experience difficulties sustaining attention and managing impulses during less stimulating activities \cite{groen2020testing}. As video-based learning and training become increasingly prevalent in higher education and workplace environments \cite{noetel2021video, long2023review}, these challenges can prevent individuals with ADHD from effectively accessing critical educational resources, exacerbating existing disparities in academic achievement and professional advancement for the neurodivergent population \cite{arnold2020long, nadeau2005career, loe2007academic}.

Prior work has been examining ways to improve designs of multimedia learning materials for more effective knowledge delivery. From a theoretical lens, multimedia learning and cognitive load theories \cite{mayer2024past, cceken2022multimedia} provided foundations for guiding high-quality educational presentations \cite{clark2023learning, cavanagh2023using}. Leveraging these theoretical frameworks, researchers have designed systems to evaluate \cite{alshaikh2024implementation, kirschner2023toward}, improve \cite{fyfield2022improving, namestovski2022framework}, and automatically create \cite{chen2024automatic} educational presentations and videos. However, such efforts have predominantly stemmed from a neurotypical perspective, despite evidence suggesting distinct behaviors \cite{mayes2020sluggish, mahak2025academic} and needs \cite{hite2021describing} of neurodivergent learners. For example, while prior work indicated that picture-in-picture (i.e., small instructor window overlaying the slides) style of video lecture can lead to better learning performance for the general student body \cite{Kokoc2020-if}, recent work found that the presenter window could sometimes be distracting for users with ADHD \cite{das2021, zhu2025character}, leading to undermined information acquisition. 

The mismatches between generic video learning design principles and ADHD-specific preferences highlight the importance of \textit{adaptive learning}---tailoring educational materials to individual learners' cognitive and attentional profiles \cite{martin2020systematic, halkiopoulos2024leveraging}. 
However, little is known about how we can adapt video learning design to fulfill the needs of individuals with ADHD. While recent research has started to explore the video watching challenges faced by viewers with ADHD \cite{adhdvideoaccess, zhu2025character}, they focused on social media videos, which differ substantially from video lectures in structure, pacing, and cognitive demand \cite{lackmann2021influence}. Key research questions around ADHD individuals' video learning experiences remain unanswered: How do the multimodal elements in video lectures, such as text density, visual complexity, instructor presence, and speech patterns, shape their learning challenges? How are their challenges reflected in behavioral signals to inform adaptive systems? What coping strategies do they adopt to overcome video learning challenges and what are their video lecture design preferences? 

To answer these questions and inspire adaptive video learning for ADHD, we conducted a retrospective think-aloud study \cite{alhadreti2018rethinking, van2003retrospective} with 16 participants with ADHD, each of whom watched two video lecture segments and reflected on their experiences and challenges after each viewing session. Meanwhile, the researchers monitored and recorded participants' behavioral data (e.g., gaze, actions), took note of notable behaviors and events, and probed participants with these events in the post-hoc reflection. As gaze behaviors provide substantial insights into people's attention, we also played back participants' gaze trajectories when they reflected on certain video segments to support memory recall. 

Our study revealed the challenges caused by inaccessible video lecture designs for learners with ADHD (e.g., overfocusing on dominant yet underexplained visuals) and how such challenges were reflected in viewers' behaviors (e.g., prolonged fixations indicating mind-wandering during understimulating video segments). Furthermore, we unpacked viewers' coping strategies when navigating inaccessible video lectures (e.g., strategically mind-wandering when understimulated to prevent complete disengagement) and uncovered the limitations of existing strategies (e.g., difficulty recollecting attention when video lacks clear signifiers for topic transitions). We also explored participants' practices for managing learning needs beyond the video watching process (e.g., using  AI to check their notes for knowledge gaps), and highlighted design opportunities for educators, adaptive learning systems, and beyond. 

In summary, our research contributes, to our knowledge, the first systematic examination of video lecture watching experiences for viewers with ADHD with behavioral evidence. Our study characterized inaccessible multimodal video lecture elements, uncovered coping strategies and concerns, and derived design implications to inform adaptive video learning systems for individuals with ADHD.

\section{Background \& Related Work}
Our work builds on prior research that highlights the challenges
faced by individuals with ADHD during video learning, technological efforts for supporting learners with ADHD, and adaptive and personalized learning methods that motivated our work. We introduce them below to contextualize our research.

\subsection{Challenges of Video Learning for Individuals with ADHD}
Attention-deficit/hyperactivity disorder (ADHD) is a neurodevelopmental disorder that affects 7.6\% of children and 6.8\% of adults \cite{Salari2023}, with inattention and/or hyperactivity/impulsivity \cite{Wilens2010-hq} being common symptoms. Exacerbated by the high rate of comorbidity (e.g., learning disabilities, dyslexia) for ADHD \cite{Sobanski2006, reale2017}, individuals with ADHD could experience significant challenges in learning, including task completion \cite{hoza2001academic, modesto2013motivation}, organization \cite{kofler2018working, bikic2017meta}, sustained focus \cite{marchetta2008sustained, tucha2017sustained}, and information acquisition from various learning materials \cite{rucklidge2002neuropsychological, kim2014visual, blomberg2021effects}, leading to higher rates of underachievement in both educational and professional contexts \cite{arnold2020long, nadeau2005career, loe2007academic}.

The increasing prevalence of video-based learning has created new opportunities to alleviate ADHD-related challenges in learning \cite{levenberg2023learning, karnad2013neurodiversity}. However, it simultaneously introduces new challenges. As a multimodal format for delivering information, video lectures often present complex visual elements---such as text, graphics, and instructor figures---alongside auditory components with varying tones, pacing, and speech content \cite{Li2022, Wittenberg2021}. While such multimodality can enrich learning \cite{noetel2021video}, it also increases cognitive and attentional demands \cite{costley2021effects}, especially for individuals with ADHD who often experience difficulties in sustaining attention and processing information efficiently \cite{kofler2018working, roberts2012constraints}. 

Prior work has shown that people with ADHD may be distracted or irritated by both visual and auditory stimuli in videos \cite{fabio2015adhd, emam2025enhancing}. For example, Jiang et al. \cite{adhdvideoaccess} conducted semi-structured interviews and found that viewers with ADHD can be overstimulated by multimodal elements such as flashing lights and abrupt sounds. Zhu et al. \cite{zhu2025character} collected and analyzed data from popular video-sharing platforms and found that viewers with ADHD can be distracted by major visual and auditory components of videos, including speakers, content layouts, and background visuals and audio. These efforts, however, focused on social media videos. No work has comprehensively and deeply examined the experiences of viewers with ADHD with respect to video lectures, which pose unique challenges due to their complex, relatively static, and information-dense presentation format \cite{choe2019student, alpert2019video}. To address this gap, our work deeply characterizes the video lecture watching experiences of individuals with ADHD, uncovering the specific design features that exacerbate or alleviate their attentional and cognitive challenges.

\subsection{Technologies to Support Learning for Individuals with ADHD}

Prior research has explored different assistive technologies to support learners with ADHD across diverse age groups. For children and adolescents with ADHD, who often learn in a more structured environment, researchers have been designing tools to help children regulate their learning behaviors and attention in schools \cite{sonne2015, wong2023effectiveness} and home-learning environments \cite{lopez2020development, 10.1145/2858036.2858157}. In contrast, adult learners with ADHD often operate in more flexible environments where they are expected to self-regulate their learning \cite{cohen2012importance}, resulting in distinct and more pronounced challenges \cite{meaux2009adhd}. Accordingly, recent work has started to explore tools for adult learners with ADHD to mitigate educational and professional challenges, including task management \cite{zhu2026scaffolding, chen2026not}, communication \cite{zhang2025understood}, and attention support \cite{riaz2024interaction}. For example, Cuber et al. \cite{cuber2024} designed a VR studying environment with noise cancellation to help college students with ADHD focus on their schoolwork. Lalwani et al. \cite{Lalwani2025} designed a social robot as a companion for college students with ADHD during academic tasks, showing that the presence of a social robot serves a body-doubling role---a common ADHD strategy to increase focus and productivity through co-presence \cite{eagle2024something}.

Given the opportunities and challenges that video-based learning affords for individuals with ADHD \cite{levenberg2023learning}, recent work has also started to explore video accessibility \cite{adhdvideoaccess, zhu2025character} for individuals with ADHD. For example, Zhu et al. \cite{focusview} designed a video customization interface that allowed viewers with ADHD to simplify a video by removing visual and auditory distractions. However, their work focused on general informational videos and did not account for the more complex and structured design of video lectures \cite{chorianopoulos2018taxonomy}, which require viewers to continuously coordinate attention across multiple elements (e.g., text, visuals, instructors). In the context of video learning, Das et al. \cite{das2025towards} created a video lecture watching system with supportive features for learners with ADHD, including video chaptering, summarization, and built-in teaching assistants. However, their work focuses on auxiliary support features and does not engage with the underlying multimodal design of video lectures, which can meaningfully influence how learners process and acquire information \cite{chen2015effects}. Complementing these auxiliary support tools, our work turns attention to the video lecture content itself, examining how its multimodal design elements shape the learning experiences of individuals with ADHD and where opportunities for adaptation may lie.

\subsection{Accessible Learning via Adaptation and Personalization}






Extensive research has recognized the diverse learning needs across individuals, shaped by differences in sensory profiles, prior knowledge, and learning contexts \cite{schmeck1988individual, alwawi2026beyond}. Adaptive learning systems have sought to address this diversity by tailoring educational content and delivery to individual learners, drawing on techniques such as learner modeling \cite{vandewaetere2011contribution, wang2025development}, intelligent tutoring \cite{phobun2010adaptive}, and machine learning-driven content recommendations \cite{chen2018recommendation, sabeima2022towards}. Such efforts to personalize learning have also extended to the accessibility domain, where researchers design systems to adapt educational materials for learners with diverse abilities \cite{alhosban2024alt, gevorgyan2024use}. For example, Batanero et al. \cite{batanero2014considering} investigated accessible e-learning platforms that offer learning materials in different modalities (e.g., text, speech, sign languages) to accommodate different needs of learners with disabilities. Standen et al. \cite{Standen2020} designed an adaptive learning system for learners with intellectual disabilities that selects appropriate learning materials based on the learner's engagement state, which was predicted using
machine learning. These systems highlighted the benefits of moving from a one-size-fits-all approach to personalized learning based on users' preferences and needs.

Nonetheless, neurodivergence---and ADHD in particular---presents unique challenges for adaptive learning. First, strategies to regulate neurotypical learners' engagement might not apply to learners with ADHD. For example, while increasing stimulation (e.g., richer visuals, dynamic animations) is a common technique for sustaining engagement \cite{zhang2025influence, mardhatilah2023digital}, such enhancements might overwhelm or distract individuals with ADHD \cite{adhdvideoaccess}. Additionally, individuals with ADHD might exhibit engagement and disengagement differently than neurotypical people \cite{Vile_Junod2006-tk}. 

Despite this complexity, efforts to personalize learning specifically for individuals with ADHD remain limited. Thawalampola et al. \cite{Thawalampola2024} focused on adapting text-based educational materials via content chunking and simplification based on the behaviors (e.g., head movement, facial expressions) of learners with ADHD. Similarly, Yadav \cite{yadav2025} proposed a framework to personalize textbook content for children with ADHD by analyzing multimodal data and neuropsychological tests. However, no work has focused on adapting video lectures to support learning for individuals with ADHD. Thus, we extend the exploration of adaptive learning for ADHD into the video lecture domain, generating empirical insights and design implications that can inspire future video adaptation systems for learners with diverse cognitive and attentional needs.

\section{Methodology}
To understand the video learning experiences of learners with ADHD, we conducted a \textit{retrospective think-aloud study} \cite{van2003retrospective, alhadreti2018rethinking} with 16 participants with ADHD. To preserve participants' natural video-learning behaviors without interrupting or distracting them, which is particularly important for ADHD learners, we asked participants to concentrate on video watching and invited them to reflect on their experience \textit{post-hoc}. To alleviate challenges with working memory for individuals with ADHD \cite{kofler2018working}, we used participants' observed behaviors (e.g., gaze trajectories, interaction logs) as probes to support their memory recall. We explain the study setup, procedure, and analysis method below. 

\subsection{Participants}
We recruited 16 participants with ADHD (P1-16, 9 females, 7 males) whose ages ranged from 20 to 51 (\textit{Mean} = 28.4, \textit{SD} = 10.0) via email lists. A participant was eligible if they were at least 18 years old and self-reported as having ADHD. Fourteen participants had been clinically diagnosed with ADHD, and two (P6, P15) were in the clinical diagnostic process after initial screening by the university health services. All participants reported substantial experiences watching video lectures. Table \ref{tab:users} provides participants’ detailed demographic information. 
Participants were compensated \$20 per hour and reimbursed for travel expenses. This study was approved by the Institutional Review Board (IRB) at our university.

\begin{table*}[h]
\footnotesize 
\centering
\caption{Demographic Information of Participants.}
\begin{tabular}
{C{0.6cm}C{0.6cm}C{1.5cm}C{3.0cm}C{2.8cm}C{4cm}}
\toprule

\textbf{PID} & \textbf{Age} & \textbf{Gender} & \textbf{Diagnostic Status} & \textbf{Field of Study/Work} & \textbf{Frequency of Watching Video Lectures} \\
\midrule 

P1 & 20 & Female & Clinically diagnosed at 16 & Actuarial Science & 30\% of school time\\ \hline
P2 & 20 & Female & Clinically diagnosed at 19 & Psychology \& Social Welfare & Weekly \\ \hline
P3 & 36 & Male & Clinically diagnosed as a child and as an adult & Metabolism & Daily \\ \hline
P4 & 50 & Female & Clinically diagnosed at 35 & Biology & Weekly \\ \hline
P5 & 28 & Female & Clinically diagnosed at 26 & International Study \& Clinical Research & Weekly \\  \hline
P6 & 20 & Female & In the diagnostic process & Microbiology &  Watch many video lectures with varied frequency \\ \hline
P7 & 22 & Female & Clinically diagnosed at 11  & Speech Pathology & Weekly \\ \hline 
P8 & 28 & Female & Clinically diagnosed at 14 & Primary Care & Daily when taking classes \\ \hline
P9 & 51 & Male & Clinically diagnosed at 28 & Physician Assistant & Daily \\ \hline
P10 & 22 & Transgender male & Clinically diagnosed at 17/18 & History \& Film Studies & Watch many video lectures with varied frequency \\ \hline
P11 & 34 & Male & Clinically diagnosed at 30/31 & Biomedical Research & Daily \\ \hline
P12 & 22 & Male & Clinically diagnosed at 10 & Mechanical Engineering & Weekly\\ \hline
P13 & 21 & Female & Clinically diagnosed at 16 & Neurobiology & Daily \\ \hline
P14 & 31 & Male & Clinically diagnosed at 17  & Journalism &  Watch many video lectures with varied frequency \\ \hline
P15 & 25 & Female & In the diagnostic process & Data Science & Most of school time\\ \hline
P16 & 24 & Male & Clinically diagnosed at 21  & Information Systems & Weekly \\ 

\hline
\end{tabular}
\label{tab:users}
\end{table*}

\begin{figure*}
    \centering
    \includegraphics[
        width=0.95\linewidth,
        alt= {Illustration of a two-stage study setup for video learning with eye tracking. (A) During video viewing, a participant sits at a desk watching a lecture video on a monitor equipped with an eye tracker and webcam. In a separate partitioned space, a researcher monitors the participant on two screens: one showing the video with real-time gaze overlays and transcript panel (labeled 1), and another showing a recording of the participant’s face alongside their screen activity (labeled 2).(B) During retrospective reflection, the participant and researcher sit together facing a screen displaying the lecture video with gaze trajectories overlaid, discussing the participant’s viewing experience.}
    ]{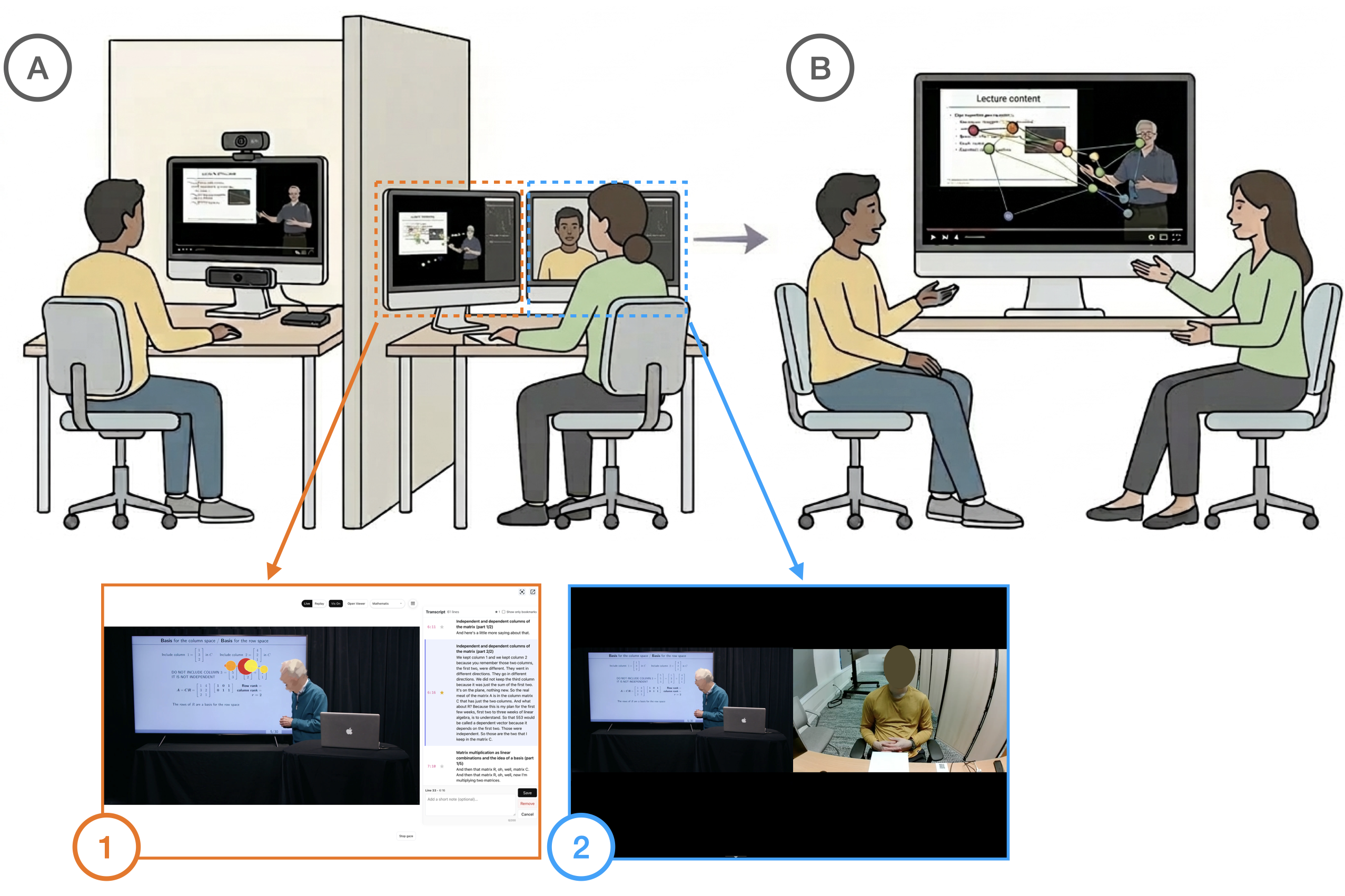}
    \caption{Our two-stage study interface. (a) During video viewing, the participant watches a video lecture on an eye-tracker-equipped display while the researcher monitors their behavior across two screens: (1) a mirror of the participant's video view with recent gaze trajectories overlaid and a synchronized transcript panel supporting bookmarking and note-taking, and (2) a screen recording of the participant's display alongside their face recording. (b) During gaze-supported reflection, the researcher advances the video to a given timestamp, and participants review their gaze trajectories overlaid on the video frame to recall and reflect on their viewing experience.}
    \Description{Illustration of a two-stage study setup for video learning with eye tracking. (A) During video viewing, a participant sits at a desk watching a lecture video on a monitor equipped with an eye tracker and webcam. In a separate partitioned space, a researcher monitors the participant on two screens: one showing the video with real-time gaze overlays and transcript panel (labeled 1), and another showing a recording of the participant’s face alongside their screen activity (labeled 2).(B) During retrospective reflection, the participant and researcher sit together facing a screen displaying the lecture video with gaze trajectories overlaid, discussing the participant’s viewing experience.}
    \label{fig:interface}
\end{figure*}

\subsection{Apparatus}

To understand participants' experiences with diverse video lecture designs, we prepared a curated set of video lectures spanning a range of content and presentation styles. We also developed a study interface supporting video watching, real-time behavior monitoring (e.g., gaze, body), and gaze trajectory playback. We describe the study apparatus below.

\subsubsection{Video Lecture Selection}
We selected eight video lectures that span a wide range of content and presentation styles (Table \ref{tab:video_list}). Drawing on prior taxonomies and categorizations of video lectures \cite{chorianopoulos2018taxonomy, lee2023lecture}, we carefully curated a list of video lectures containing diverse designs, including the presence of the instructor and the slides, the usage and style of visuals (e.g., text, images, data visualizations, tables, and formulas), and the style (e.g., pace, tone) of the speech. We trimmed each selected lecture to 13 to 17 minutes with natural segment boundaries, informed by prior work on effective video lesson length \cite{yu2022effects}. We chose this duration range to reduce participants' fatigue while preserving enough length for the attentional challenges of video learning to emerge \cite{manasrah2021short}.

\begin{table*}[h]
\scriptsize
\setlength{\tabcolsep}{3pt}
\centering
\caption{Overview of selected video lectures.}
\begin{tabular}{C{0.5cm} C{1.2cm} C{3.5cm} C{2.5cm} C{1.2cm} C{1.4cm} C{5.8cm}}
\toprule
\textbf{VID} & \textbf{Preview} & \textbf{Title} & \textbf{Channel} & \textbf{Topic} & \textbf{Segment} & \textbf{Design Characteristics} \\
\midrule
 V1 & \includegraphics[width=1.0cm, alt= {A lecture video showing an instructor standing beside a large display with mathematical equations, gesturing while speaking, with a laptop on a table.}]{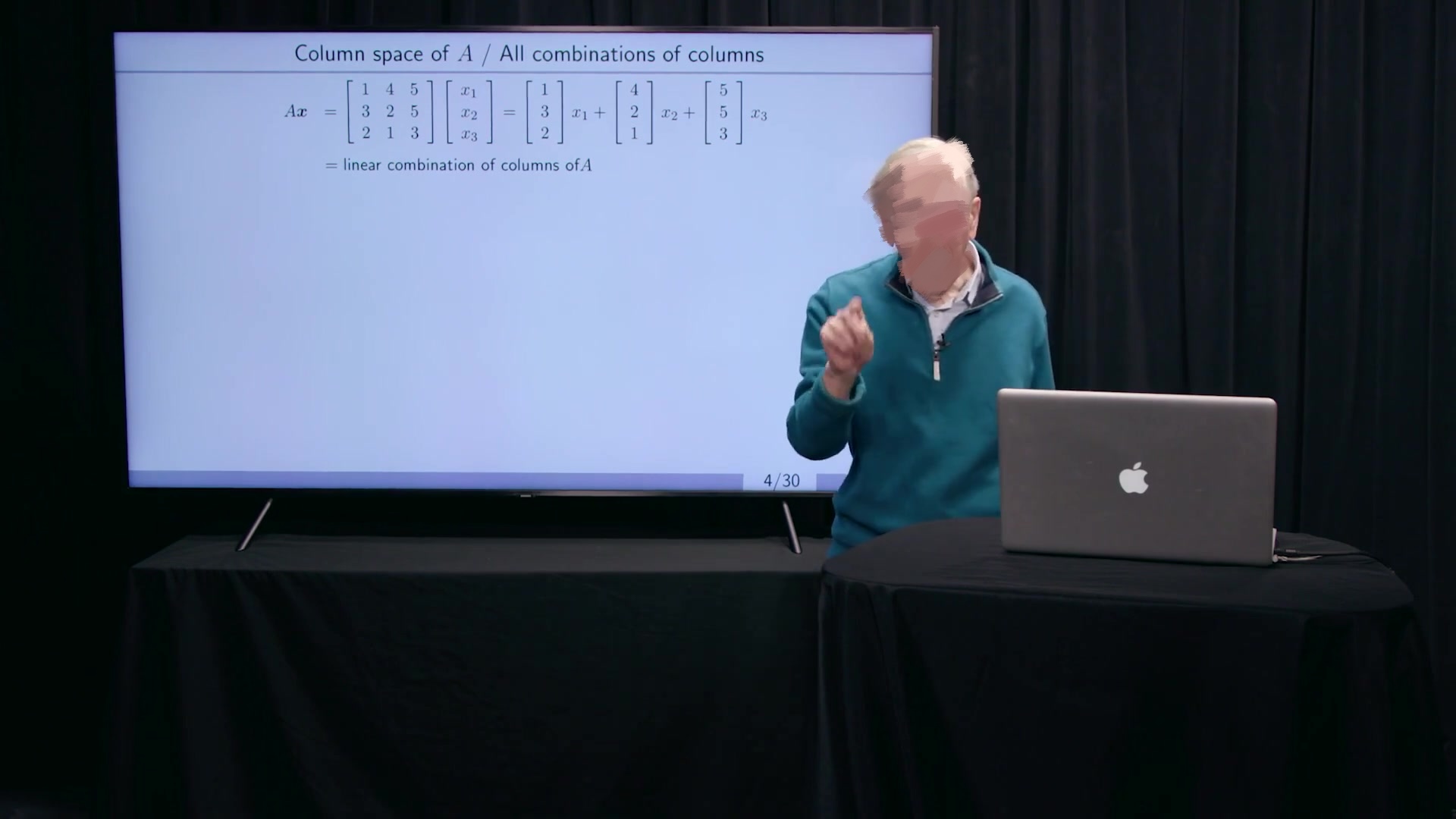} \Description{A lecture video showing an instructor standing beside a large display with mathematical equations, gesturing while speaking, with a laptop on a table.} & 
\href{https://www.youtube.com/watch?v=azzrfdysfI0} {The Column Space of a Matrix}  & MIT OpenCourseWare  & Mathematics & 0:00--13:59 & switching view, on-screen instructor, real screen, text-only slides  \\
 V2 & \includegraphics[width=1.0cm, alt= {A slide featuring an image of an octopus labeled "Octopus," alongside a sidebar with lecture interface elements and a small inset video of the instructor.}]{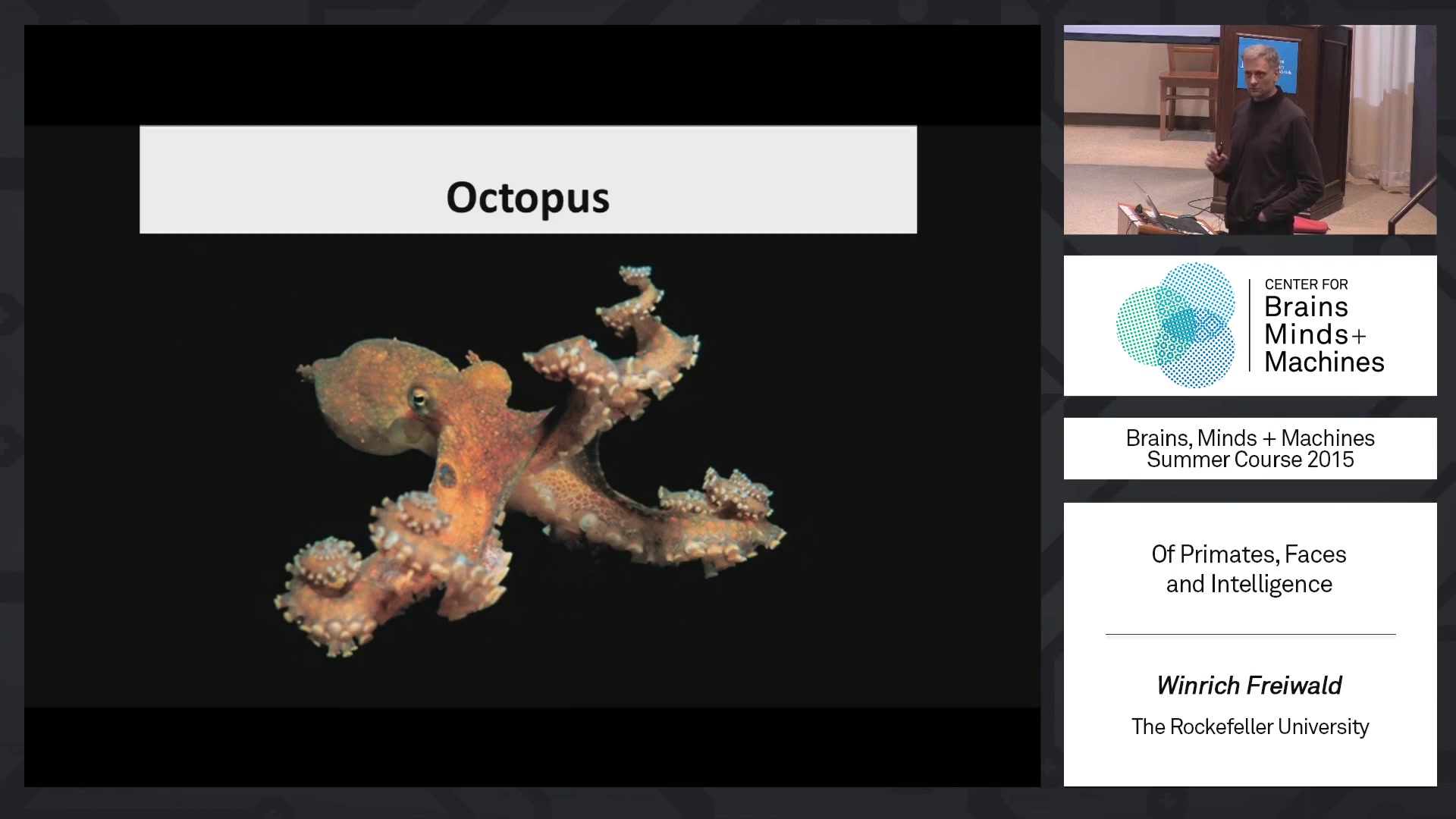} \Description{A slide featuring an image of an octopus labeled "Octopus," alongside a sidebar with lecture interface elements and a small inset video of the instructor.} & \href{https://www.youtube.com/watch?v=8PcPpVQK7N8} {Primates, Faces, \& Intelligence}  & MIT OpenCourseWare & Neuroscience & 0:00--15:45 & static view, picture-in-picture (PiP) instructor,  text- and image-dense slides, misc information, high speech rate \\
 V3 & \includegraphics[width=1.0cm, alt= {A lecture slide titled "Widening Inequalities of Place" with a bullet-point outline, presented next to an instructor standing and speaking.}]{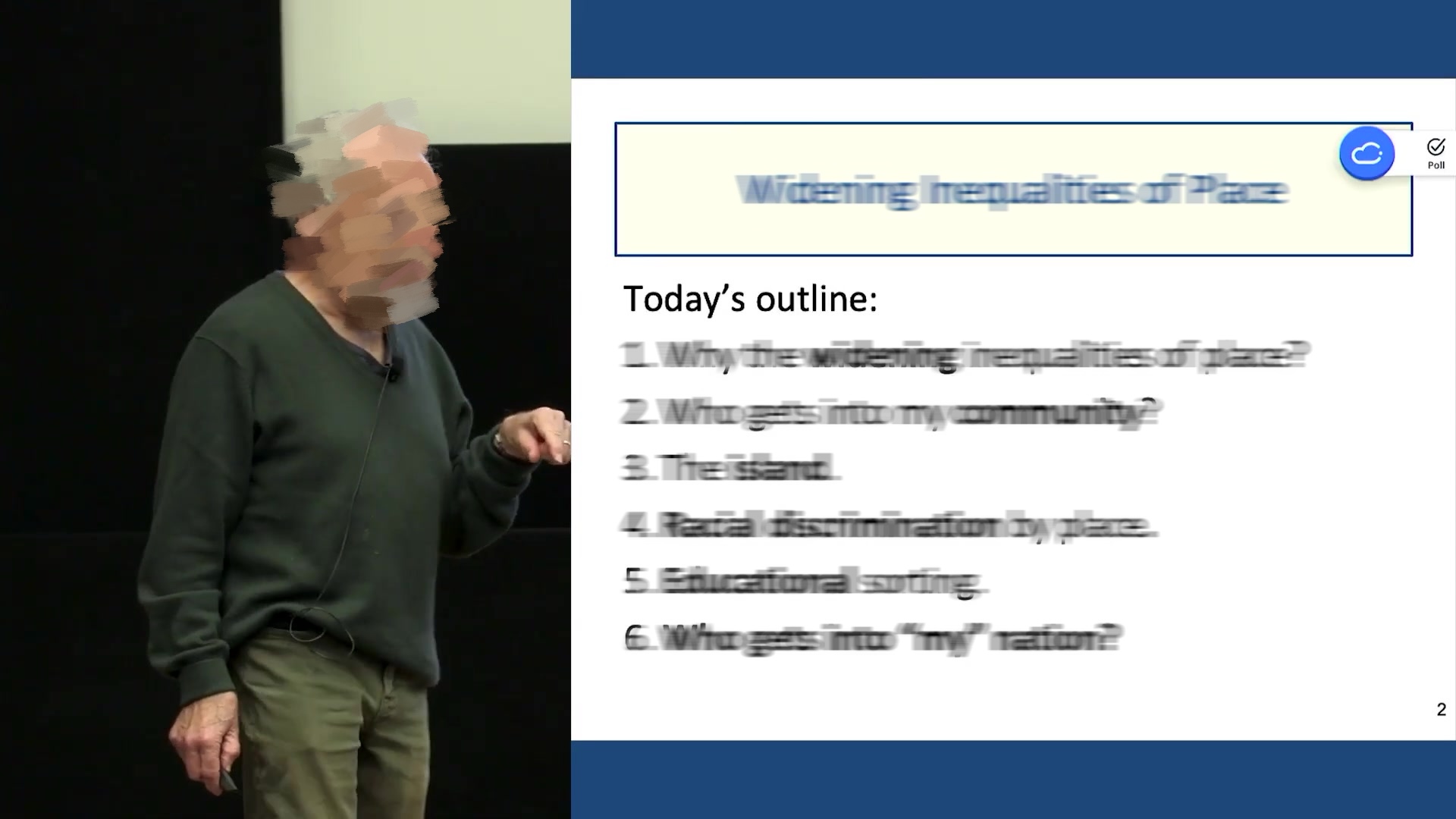} \Description{A lecture slide titled "Widening Inequalities of Place" with a bullet-point outline, presented next to an instructor standing and speaking.} & \href{https://www.youtube.com/watch?v=owqQQvmewaY&list=PLOLArO56vjuoeaIPzKQibBDbx2m_Rfsit&index=4}{``Widening Inequalities of Place''} & Robert \& Reich & Public Policy & 3:06 -- 17:28  &  switching view, on-screen instructor, slides overlay, interactive activities, low speech rate \\
 V4 & \includegraphics[width=1.0cm, alt={A classroom lecture scene with an instructor standing at a podium in front of a wooden wall, speaking toward an audience.}]{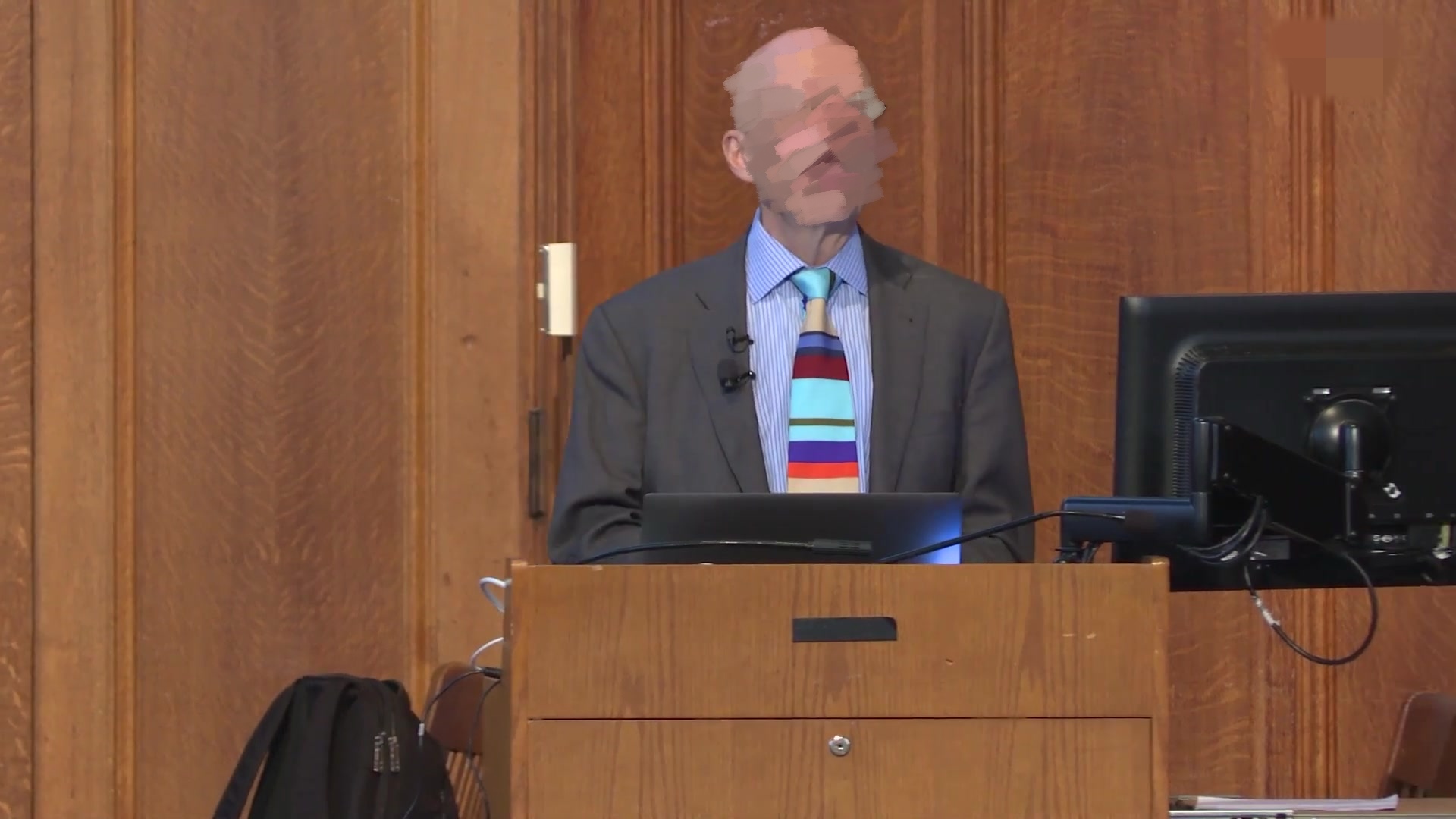} \Description{A classroom lecture scene with an instructor standing at a podium in front of a wooden wall, speaking toward an audience.} & \href{https://www.youtube.com/watch?v=9dULs7w8b-0} {Privatizing Government I: Utilities, Eminent Domain, and Local Government} & YaleCourses & Political Science & 0:00--15:29& switching view, on-screen instructor, diagram-dense slides, talking-head-only segments\\
 V5 & \includegraphics[width=1.0cm, alt={A slide titled “In vitro” displaying diagrams, charts, and image thumbnails related to scientific or computational content.}]{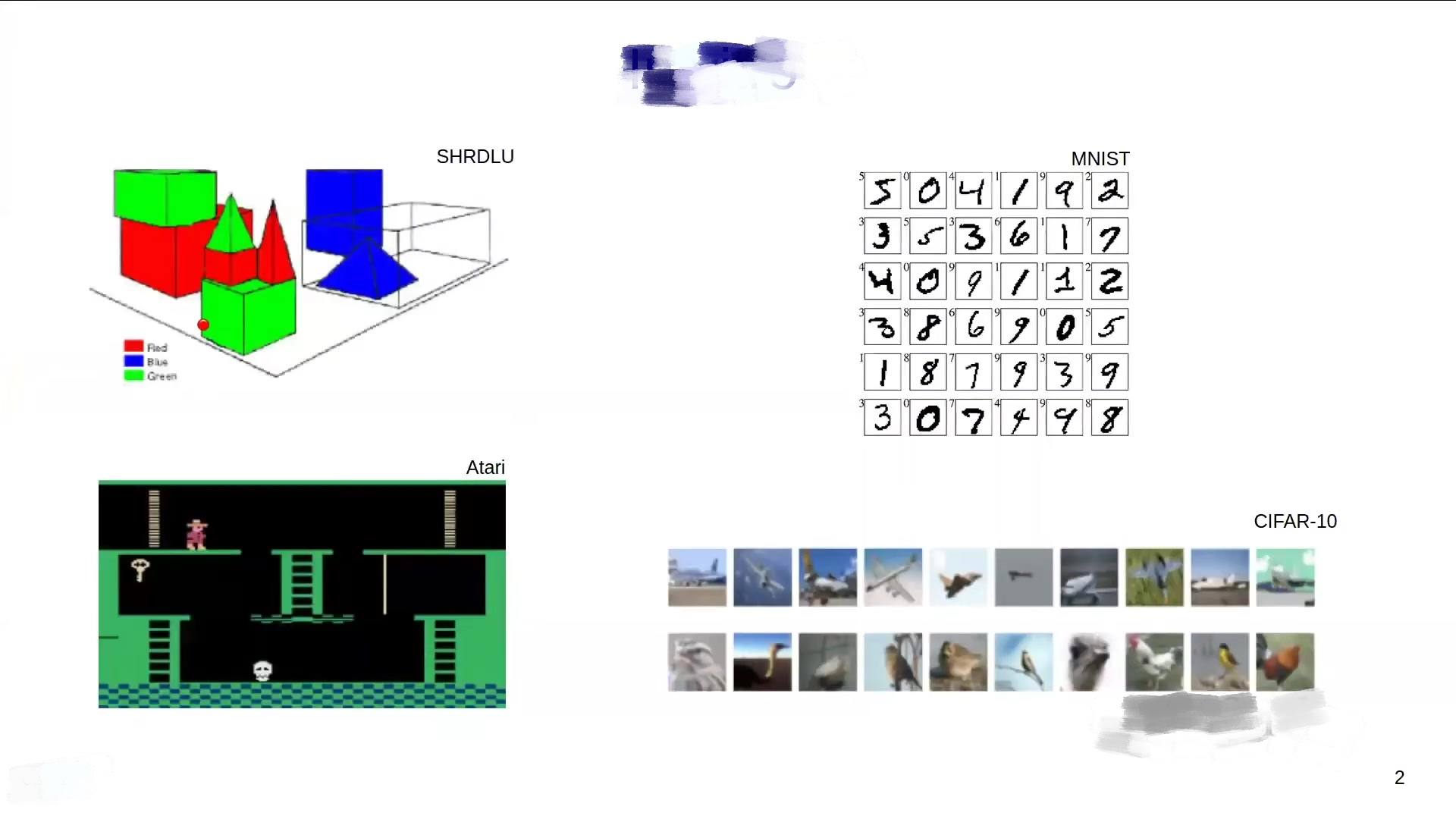} \Description{A slide titled “In vitro” displaying diagrams, charts, and image thumbnails related to scientific or computational content.} & 
\href{https://www.youtube.com/watch?v=C0IhR4D5KYc} {Artificial Intelligence Today} & Stanford Online & Computer Science & 0:00 -- 13:45 & static view, no instructor, image-dense slides \\
 V6 & \includegraphics[width=1.0cm, alt={A lecture video showing an instructor gesturing toward a slide with a graph labeled "Risk of disease varies with both genotype and environment."}]{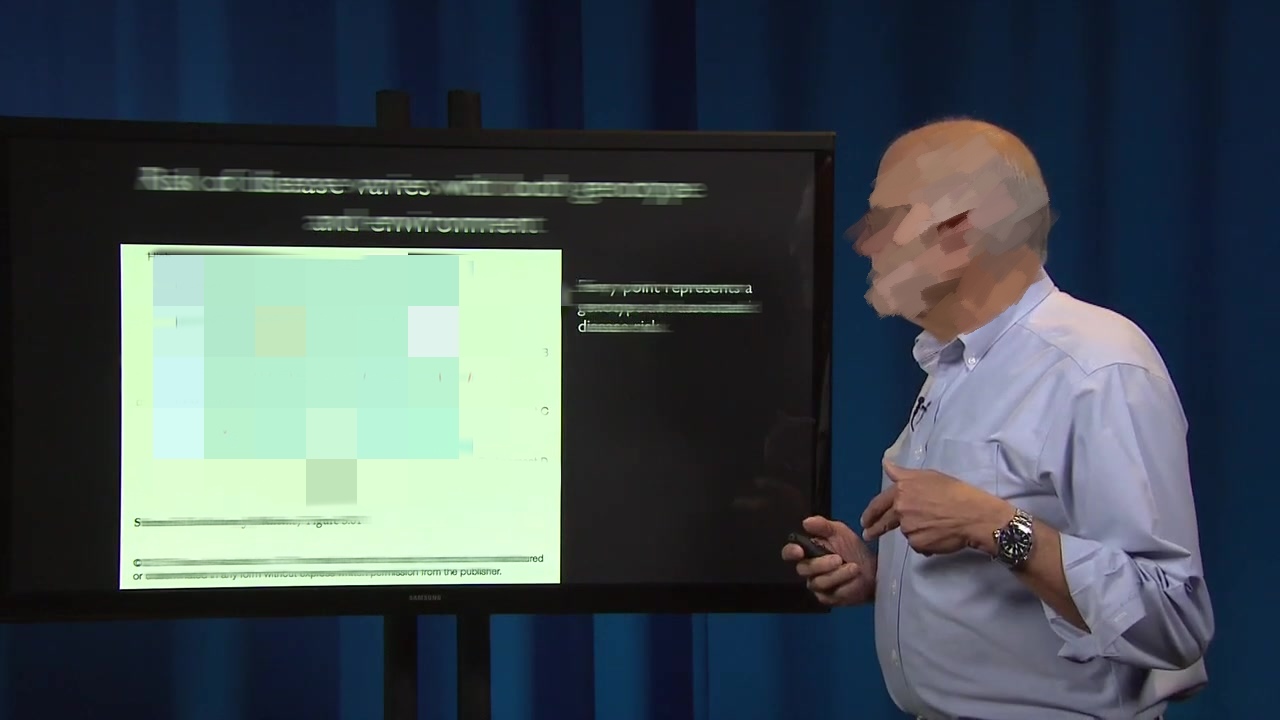} \Description{A lecture video showing an instructor gesturing toward a slide with a graph labeled "Risk of disease varies with both genotype and environment."} & \href{https://www.youtube.com/watch?v=MqU4b_VZNoo} {What is a disease? Introduction} & YaleCourses & Health & 0:00--16:32 & static view, on-screen instructor, real screen, text-dense slides, slide-reading, low speech rate\\
 V7 & \includegraphics[width=1.0cm, alt={A slide titled “Problems in Prehistoric Art” showing illustrated panels and accompanying explanatory text.}]{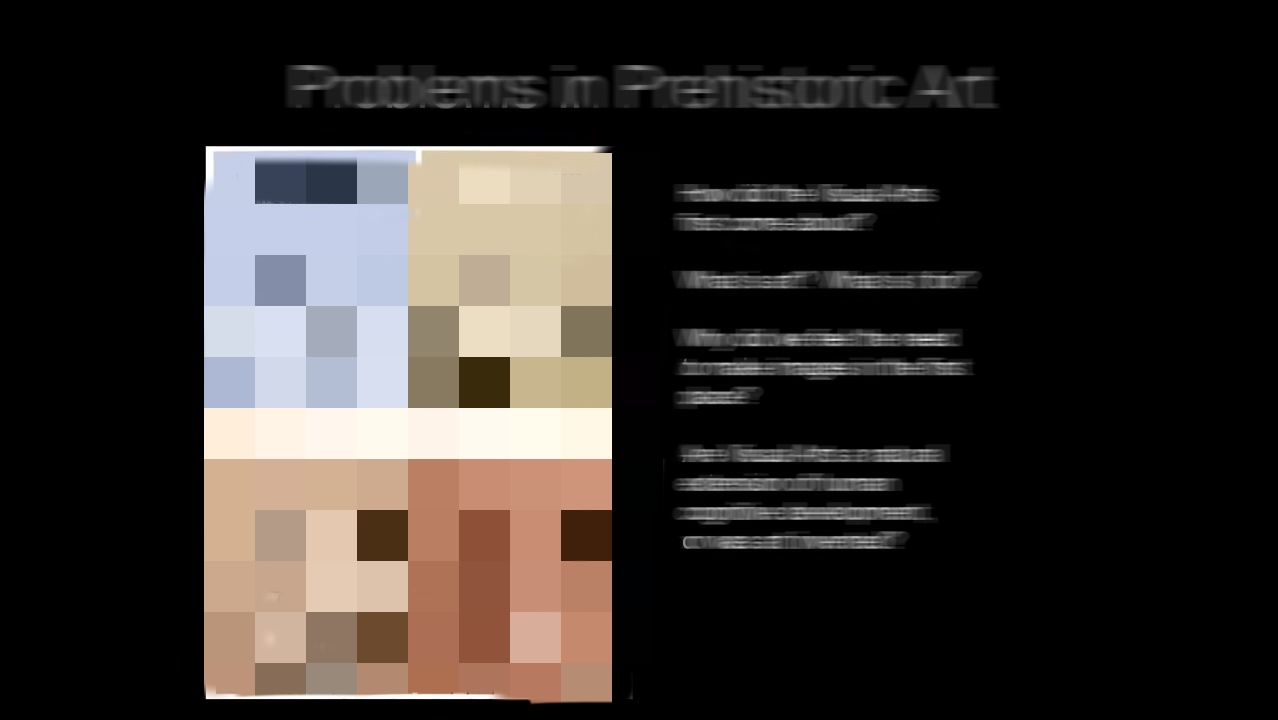} \Description{A slide titled “Problems in Prehistoric Art” showing illustrated panels and accompanying explanatory text.} & \href{https://www.youtube.com/watch?v=P_tkKoXDfdg} {Prehistoric Art}  & 
Art History with Travis Lee Clark  & Art History & 0:43--17:20
 &  static view, no instructor, image-dense slides, variation in tone\\
 V8 & \includegraphics[width=1.0cm, alt={A lecture slide titled “Payment Systems” with text on the left and a small inset video of an instructor speaking in the corner.}]{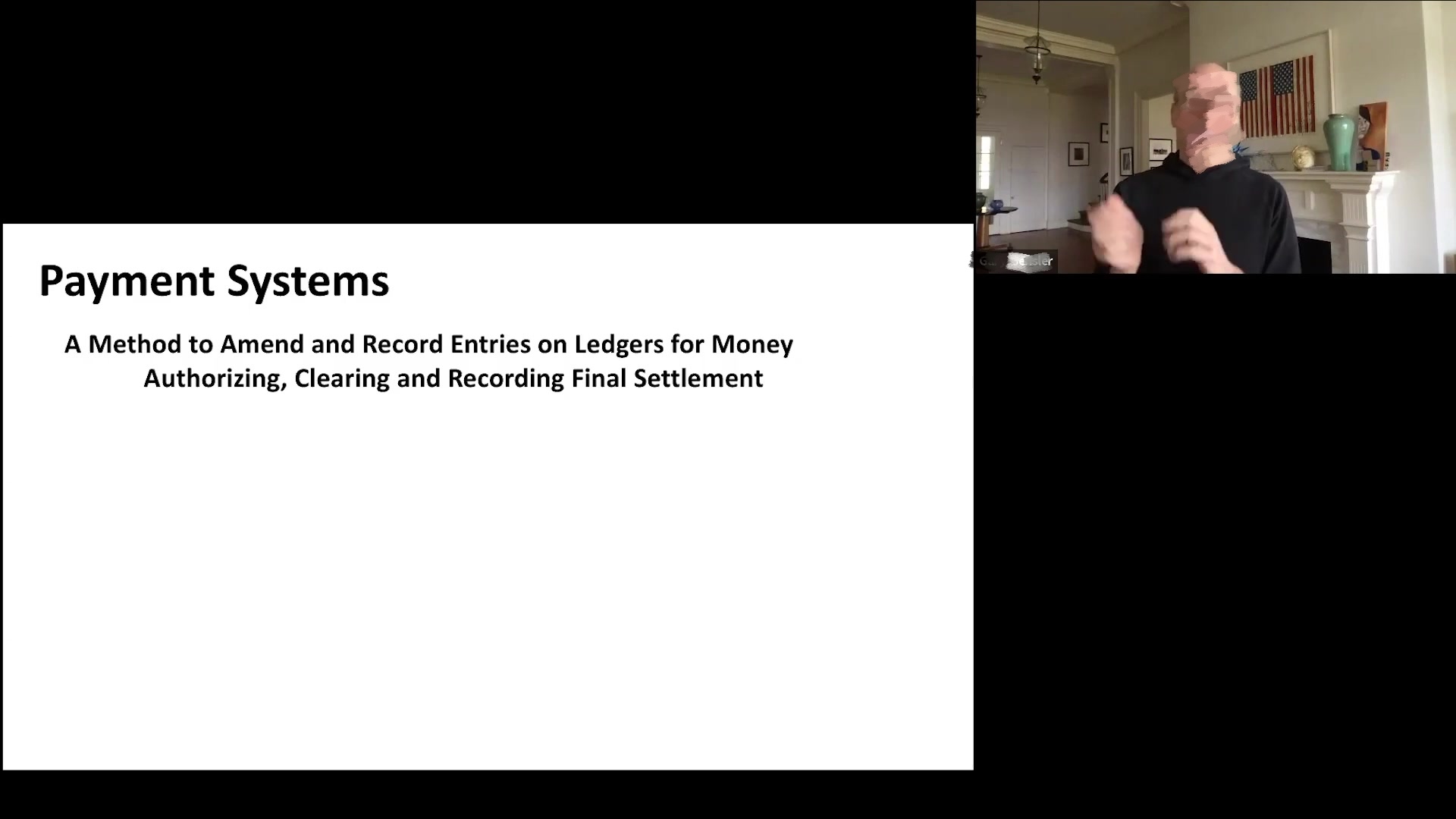} \Description{A lecture slide titled “Payment Systems” with text on the left and a small inset video of an instructor speaking in the corner.} & \href{https://www.youtube.com/watch?v=4FGNLl9Btfw} {Payments} & MIT OpenCourseWare & Business  & 19:18--35:51  & static view, PiP instructor, diagram-dense slides, interactive activities\\
\bottomrule
\end{tabular}
\label{tab:video_list}
\end{table*}

\textbf{\textit{Video Assignment.}} To simulate realistic learning experiences, where adult learners typically have some agency in choosing what they learn \cite{galotti2019students}, we used a preference-aware assignment procedure. Prior to the study, participants indicated video lecture topics they would prefer to \textit{avoid}, and videos were randomly assigned from the remaining pool, with each video viewed by four participants in a counterbalanced order. This procedure also helps mitigate the attention decline caused by lack of interest in the videos.


\subsubsection{Study Interface}

We designed a two-stage interface (Figure \ref{fig:interface}) to support (a) naturalistic video viewing with concurrent behavioral monitoring, and (b) gaze-supported retrospective reflection. 

\textbf{\textit{Video Viewing and Behavior Monitoring.}} 
Participants viewed the video lecture using a 24-inch display (1920$\times$1200) with a Tobii Pro Fusion (120 Hz) eye tracker mounted at the bottom. They viewed the video using a standard HTML video player (Figure \ref{fig:interface}A) with the common controls of mainstream video learning platforms: play, pause, seek, playback speed adjustment, and caption toggle.  To minimize distractions during viewing, the researcher was seated in a separate physical space and monitored participants' behavior across two screens. The first screen (Figure \ref{fig:interface}A(1)) mirrored the participant's video view with gaze trajectories overlaid, indicating where the participant was looking in real time. It also included a synchronized transcript panel supporting researcher's bookmarking and note-taking of notable behaviors. The second screen (Figure \ref{fig:interface}A(2)) displayed a recording of the participant's computer screen alongside the participant's real-time video recording through a webcam, allowing the researcher to observe their interactions with the video and physical behaviors during video learning.

\textbf{\textit{Gaze-supported Reflection.}} 
After the viewing session, participants engaged in a retrospective think-aloud session supported by a gaze playback interface (Figure \ref{fig:interface}B). The interface consisted of a video player with visualizations of gaze trajectories. The researcher controlled the playback by advancing the video to a certain timestamp, so that participants can revisit a video segment with their gaze trajectory overlaid on the video frame. 

\subsubsection{Implementation} The study interface was implemented in React \cite{React}. Gaze data was captured via the Tobii Pro SDK in Python and processed by a Flask-SocketIO server that handled bidirectional communication between the eye tracker and the interface. Gaze trajectories were derived from raw gaze samples using a dispersion-based real-time fixation detection algorithm \cite{kumar2008improving} suitable for low-latency streaming. To ensure that the gaze visualizations were relevant to the current video content, we segmented the video into scenes (i.e., individual slides or speaker views) using an automatic scene detection tool \cite{pyscenedetect}, followed by manual verification and correction. Gaze visualizations were reset at each scene boundary, preventing carryover of gaze data across different scenes.

\subsection{Procedure}
The study consisted of a single two-hour session with four phases: 

\textbf{\textit{Initial Interview.}} We started with an initial interview covering participants’ demographics, ADHD background, experiences with watching video lectures, and general challenges and coping strategies that they encountered and adopted during video learning. We then informed participants about behavioral data collection (i.e., gaze, face recordings, interaction events), and conducted gaze calibration using a 14-dot calibration and 5-dot validation \cite{wang2025characterizing}. 

\textbf{\textit{Video Learning.}} After calibration, participants were introduced to the video viewing interface, and familiarized themselves with all the video player features with a tutorial video. They then completed \textbf{two} video learning tasks. To simulate realistic video learning, participants were instructed to watch the videos as they normally would in a natural learning context, with full control over the video playback and the activities they want to perform (e.g., note taking) during video watching. Meanwhile, in a separate physical space, the researcher monitored participants’ behavioral data in real time, and took notes of any notable behaviors (e.g., unexpected gaze behaviors, frequent pauses or rewinds). After each video learning task, participants completed two multiple-choice quiz questions designed by the research team to assess factual recall of the video content, encouraging them to engage with the learning tasks.

\textbf{\textit{Retrospective Think-aloud.}} Immediately following each learning task, we conducted the retrospective think-aloud session to understand participants' experiences. Participants first provided overall evaluations of the video---including familiarity and interest in the video content, perceived effectiveness of video design, and perceived content absorption---on a 7-point Likert scale (7 representing the most positive). The researcher then replayed the video, advancing through it slide by slide, and participants commented on the video design while referencing their own gaze behaviors (e.g., very long fixations) when applicable. This allowed participants' reflections to surface without being biased by the researcher's observations. Within each slide, if the researcher noted any behaviors (e.g., notable pauses, rewinds, or distinctive gaze patterns) that participants did not spontaneously address, the researcher followed up with targeted questions to complement participants' recall. 

\textbf{Exit Interview.} We concluded the study with a semi-structured exit interview to discuss whether their experiences and preferences can be generalized to other video lectures, and how they would like to improve the video lecture design. To facilitate discussion, we referred to challenges they encountered when watching the selected videos, and encouraged them to think about how to make improvements without technological limitations.

\subsection{Analysis}
We analyzed the study data using both quantitative and qualitative methods.

\subsubsection{Quantitative Analysis} We reported descriptive statistics (i.e., mean, standard deviation) of participants' self-reported Likert ratings on familiarity, interest, perceived effectiveness of video design, and perceived content absorption, as well as their post-viewing quiz accuracy. To further examine how participants' perceived content absorption related to their experiences with the video, we computed Spearman's rank correlations with Holm-Bonferroni correction \cite{zar2005spearman} between perceived content absorption and three variables: perceived video design effectiveness, interest, and familiarity. Additionally, we also reported descriptive statistics on participants' behaviors during video watching (e.g., number of pauses and rewinds). 

\subsubsection{Qualitative Analysis}

We audio-recorded all sessions and transcribed interviews using an automatic transcription service. To deeply understand participants' experiences and how they manifested in behavioral data, we aligned multiple data streams for each video learning task, including the original video, gaze trajectories, face recordings, and corresponding interview excerpts, and referenced them jointly while coding.

We analyzed the data using thematic analysis \cite{braun2006using}, jointly considering both behavioral signals (e.g., gaze patterns, off-screen behaviors, interaction events) and participants’ verbal feedback during the retrospective think-aloud session. Two researchers independently open-coded four shared samples (25\% of data), developing an initial codebook through iterative discussion that captured both observable behaviors (e.g., prolonged fixations) and experiences (e.g., mind-wandering). The researchers then divided the remaining data and continued coding independently using the shared codebook, while periodically cross-checking and discussing discrepancies to ensure consistency. New codes were added to the codebook after the researchers reached an agreement.

\section{Findings}


Participants shared diverse learning experiences with the video lectures. With moderate familiarity (\textit{Mean} = 3.72, \textit{SD} = 1.67) and moderate-to-high interest in the content (\textit{Mean} = 4.75, \textit{SD} = 1.76), participants achieved an average accuracy of 92.2\% (\textit{SD} = 18.4\%) in the post video-learning quizzes, indicating serious engagement with the learning task. Nonetheless, we observed that participants' perceived level of content absorption differed tremendously across videos (\textit{Mean} = 4.32, \textit{SD} = 1.42, ranging from 2 to 7), with video design being a critical factor. We observed a strong correlation between perceived video design effectiveness and content absorption ($\rho$ = 0.75, $p$ < 0.001), while interest ($\rho$ = 0.51, $p$ < 0.01) and familiarity ($\rho$ = 0.42, $p$ = 0.016) showed only moderate correlations with content absorption. This pattern was vividly illustrated by the contrast between two lectures: while learners of the public policy lecture (V3) and the neuroscience lecture (V2) reported comparable familiarity (\textit{Mean} = 3.00 vs. 2.75) and interest (\textit{Mean} = 5.75 vs. 5.50), their perceived absorption differed sharply (\textit{Mean} = 6.25 vs. 3.75), mirroring the gap in their ratings of video design effectiveness (\textit{Mean} = 6.75 vs. 3.25). 
P11 shared his frustration after watching the neuroscience video (V2), highlighting this as a typical example of a video lecture with inaccessible designs:

\begin{quote}
    \textit{``The most typical thing I hate is that the topic interests me, but the presentation is annoying and not adapted to the way I learn. So I'm like, I really wanted to listen to you, but I will not.''} 
\end{quote}


This finding highlights the importance of making video lecture designs accessible and adaptive to the needs of learners with ADHD. In the following sections, we unpack the video viewing challenges and coping strategies that individuals with ADHD encounter during video learning with behavioral evidence. 
\subsection{Cognitive Overload from Overwhelming Content} \label{cognitive-overload}We found that cognitive overload---arising from both overwhelming multimodal presentation 
and overwhelming amounts of information---constitutes a significant barrier to effective video learning for individuals with ADHD. In the following, we elaborate on the specific features of video lecture design that contribute to this sense of overwhelm, examine how participants’ behaviors reflect this challenge, and discuss the coping strategies adopted by participants to mitigate such challenges along with  their limitations. 

\subsubsection{Overwhelming Presentation.} \label{overwhelming-presentaiton}Participants described four types of overwhelming multimodal designs in video lecture presentation: (1) overly dense visuals, (2) unexplained visuals, (3) irrelevant visuals, and (4) low quality audio. 

\textbf{\textit{Overly Dense Visuals.}} Prior work has identified dense visuals to be cognitively overwhelming for individuals with ADHD for both reading \cite{jacobson2011working} and video-watching scenarios \cite{adhdvideoaccess}. Extending this line of work, we unveiled the unique challenges with video lectures, which contain dense information from both visual and auditory channels. We found that visual density itself was a common source of overload. Ten participants (P1-2, P4, P8-13, P16) criticized six videos (V1-4, V6, V8) as \textit{``text heavy''}, and two (P13-14) described two videos (V2, V7) as having \textit{``too many pictures'' (P13).} 


Beyond visual density alone, participants also experienced challenges with aligning instructor speech with dense visual presentation (P6, P9, P11-13). For P13, the text-heavy slides in the health lecture (V6) split her attention between the instructor's speech and the slides to the point that she could not even tell whether the speaker was reading the slides or not. P11 shared a similar experience with the neuroscience lecture (V2), highlighting how text-heavy slides caused him to lose focus:

\begin{quote}
    \textit{``As soon as he switched to the slides with all the text... I probably missed more than half of the thing he said just trying to figure out what's on the slide and what he's talking about... It's like an atomic crash.''}
\end{quote}

\begin{figure*}
    \centering
    \includegraphics[width=0.95\linewidth, alt = {Composite figure showing seven examples of gaze trajectories overlaid on lecture video frames, illustrating how participants with ADHD experienced cognitive overload and applied coping strategies. Each subfigure displays fixations as circles (larger circles indicate longer fixation duration) and saccades as connecting lines, with color transitioning from red (earlier) to yellow (later) to show temporal progression. Top row: (a) Misaligned reading, where gaze lags behind the narration while reading text; (b) Content skipping, where gaze jumps across text due to high density; (c) Mindless reading, where gaze follows text without comprehension; and (d) Overfocusing, where gaze concentrates heavily on a single visual element while trying to interpret it. Bottom row: (e) Focusing on irrelevant objects, where gaze shifts to non-essential elements such as the audience; (f) Following signifiers, where gaze tracks salient cues such as instructor gestures; and (g) Anchoring attention, where gaze fixates on a stable element like the instructor to maintain focus.}]{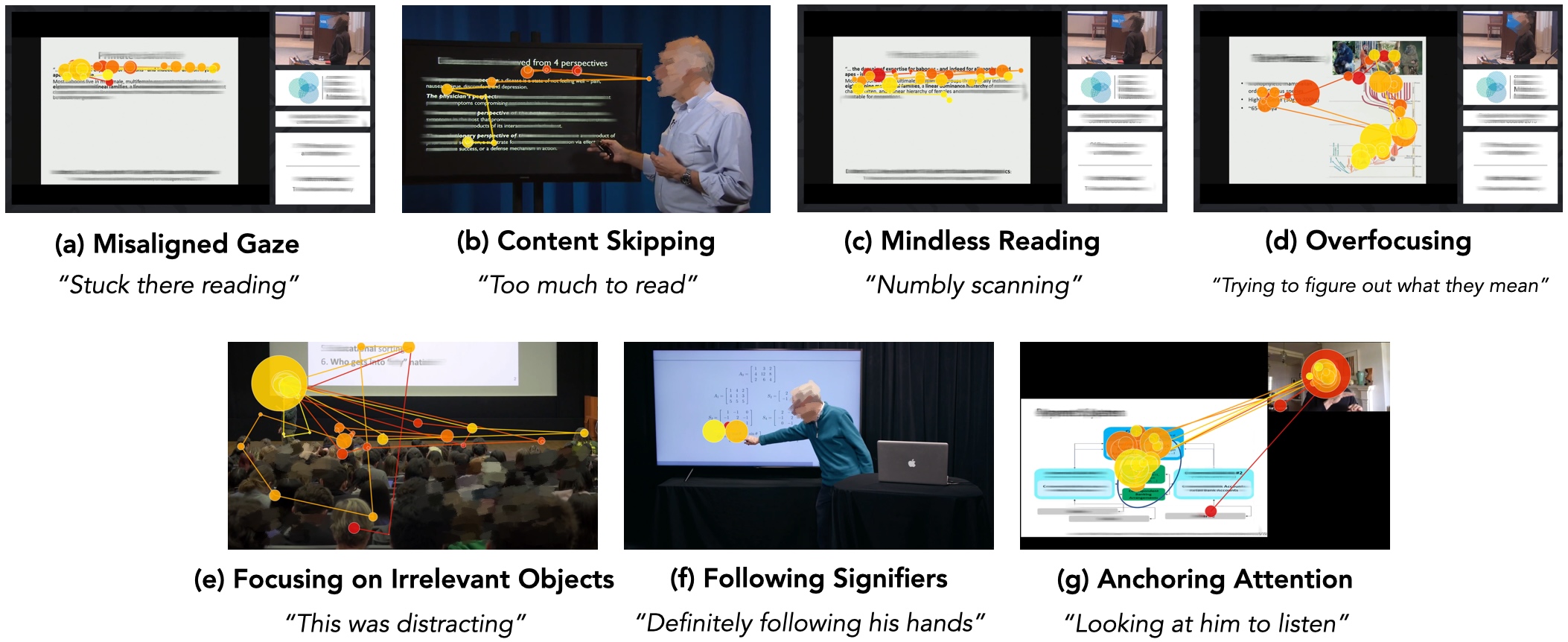}
    \caption{Examples of gaze trajectories when participants experienced \textbf{cognitive overload} and coped with these challenges. Gaze trajectory overlays on the video frames are presented as a sequence of fixations (circles) and saccades (line segments). The gaze trajectory is color-coded to show progression, transitioning gradually from red (starting point) to yellow (endpoint). The size of circle represents fixation duration. The quote underneath each frame demonstrates participants' description of their gaze trajectory with respect to their video watching experience.}
    \Description{Composite figure showing seven examples of gaze trajectories overlaid on lecture video frames, illustrating how participants with ADHD experienced cognitive overload and applied coping strategies. Each subfigure displays fixations as circles (larger circles indicate longer fixation duration) and saccades as connecting lines, with color transitioning from red (earlier) to yellow (later) to show temporal progression. Top row: (a) Misaligned reading, where gaze lags behind the narration while reading text; (b) Content skipping, where gaze jumps across text due to high density; (c) Mindless reading, where gaze follows text without comprehension; and (d) Overfocusing, where gaze concentrates heavily on a single visual element while trying to interpret it. Bottom row: (e) Focusing on irrelevant objects, where gaze shifts to non-essential elements such as the audience; (f) Following signifiers, where gaze tracks salient cues such as instructor gestures; and (g) Anchoring attention, where gaze fixates on a stable element like the instructor to maintain focus.}
    \label{fig:gaze-challenges}
\end{figure*}

Participants' gaze behaviors reflected their challenges with overly dense visuals, surfacing three distinct gaze patterns (Figure \ref{fig:gaze-challenges}a-c): (1) \textit{Misaligned Gaze} (P4, P7, P13), where participants' attention lagged behind the narration as they tried to catch up with the on-screen text. For example, P13 described her gaze trajectories in Figure \ref{fig:gaze-challenges}a as \textit{``stuck there reading,''} as she kept fixating on the previous paragraph after the instructor had moved on; (2) \textit{Content Skipping} (P1, P8-9, P11, P13, P16), where participants skimmed only a few words or phrases while skipping most of the content. For example, Figure \ref{fig:gaze-challenges}b showed P8 skipping most of the text on a slide as there was \textit{“too much to read”}; and (3) \textit{Mindless Reading} (P1, P6, P8, P11, P13), where participants' gaze appeared to be traversing lines of text but without meaningful comprehension. As shown in Figure \ref{fig:gaze-challenges}c, while P11's gaze resembled a reading behavior, he reported simply \textit{“numbly scanning”} the text. This pattern highlighted how gaze trajectory alone can disguise disengagement. As P11 commented on his gaze in Figure \ref{fig:gaze-challenges}c: \textit{``Don't be fooled. I didn't read the text.''}

While dense visuals brought challenges in general, we found that different participants could have distinct preferences for visual density. In particular, four participants (P3, P10, P13, P16) preferred dense slides despite the overload risk, viewing them as a fallback when speakers were ineffective: \textit{``Having less text is only helpful if it's a good speaker... If you miss something that was said that was important, if the slides are denser, you have something to reference to, rather than being completely lost and clueless'' (P10).} Additionally, P12 also highlighted how learners' tolerance of dense visuals could depend on their familiarity with the content: \textit{``I know what [the content] is and what I care about. But to someone who's never seen this before, they'd probably be intimidated.''}  


\textbf{\textit{Unexplained Visuals.}} Five participants (P8-11, P13) described three videos (V2, V6, V8) as having misaligned visuals and speech, where visuals signaled importance---either by containing critical information (e.g., data) or by dominating the screen (e.g., large infographics)---yet received insufficient explanation from the speaker. As a result, participants had to make excessive effort to interpret the visuals on their own, causing them to lose track of the speech content.  
For example, P13 shared her frustration when trying to interpret an infographic about the evolution of primates in the neuroscience lecture (V2). Although the image was used only for high-level illustration, it occupied a large portion of the screen and drew her attention away from the speech: \textit{``I was only looking at [the infographic], trying to figure out what they mean and how they're relevant... It would be fine if [the speaker] was gonna address the information, but he did not say anything about it.''} This challenge was manifested in her gaze data (Figure \ref{fig:gaze-challenges}d): she spent 25.2 seconds examining the infographic
, missing a substantial portion of the speech. As she reflected: \textit{``I could not even tell you what happened on [the left side of] the screen.''} 


\textbf{\textit{Irrelevant Visuals.}} Ten participants (P2, P5-6, P8-9, P11-15) highlighted the challenge of irrelevant visuals in four video lectures (V2-3, V6, V8). Such visuals are characterized as \textit{``not helpful'' (P12)} and \textit{``distracting'' (P2)} for content understanding, including images that did not offer additional information (e.g., a photo of an obese individual when discussing homeostasis-related diseases; P2, P6, P8-9, P11-13), views of lecture halls and audience (P5, P14-15), decorative images (P8, P11), and information about the speaker and educational institution (P11). When interacting with irrelevant visuals, participants' gaze would attend to those visuals rather than the main lecture content, sometimes with task-irrelevant thoughts. For example, P5 commented on her highly-dispersed gaze trajectories that explored the audience (Figure \ref{fig:gaze-challenges}e) when the public policy video (V3) switched to a view containing the audience: \textit{``This was distracting... I was definitely trying to read [the audience]'s shirt.''} 

Participants' opinions differed in what they considered irrelevant and distracting. For example, while seven participants (P2, P6, P8-10, P12, P14) found purely illustrative images distracting, four participants (P4, P13, P15-16) appreciated them as memory aids: \textit{``[The image] helps me remember what this slide is about---it's like a visual anecdote'' (P15).}


\textbf{\textit{Low-quality Audio.}} Finally, participants (P5, P9, P12) noted that background noise and disfluent speech patterns can be distracting: \textit{``The way some people talk... the quality of the recording... I'd rather you just give me a transcript and let me read it for myself'' (P9).}


\subsubsection{Overwhelming Information}

In contrast to prior findings \cite{zhu2025character} that highlighted individuals with ADHD's preferences for faster-paced content, we found that fast and dense information for video learning could lead to the challenge of information overload, with five participants (P2, P8-9, P11, P13) complaining that a video was \textit{``too much''} (P11) for four videos (V1-2, V6, V8). P9 shared the overstimulated experience with the video discussing the monetary system in the United States (V8):

\begin{quote}
    \textit{``I'm not interested anymore. [The instructor] don't view me as a human. They view me as a bucket. They're just dumping information into me, and that's dumb---it's like drinking from a fire hose'' (P9).}
\end{quote}

This feeling of overwhelm had led P9 to exhibit disengagement from the video. After he sped up the video to 1.25$\times$ due to \textit{``loss of interest''}, he spent 67.6\% of time looking away from the screen, compared to 36.0\% prior to speed adjustment. Similarly, P8 \textit{``gave up at a point''} when watching the health video (V6) as she \textit{``couldn't absorb any [content] anymore.''} As a result, she \textit{``took a break''} by looking away from the video for 52.1 seconds. P8 attributed this overwhelm to the cumulative burden of holding multiple unrelated contexts in mind at once, each introduced by a different point on the slide: \textit{``All the points [on the slide] talk about different information... You had to figure out and remember the context for each point, otherwise you're going to lose track... And this happens [throughout the video] over and over again.''}


\subsubsection{Coping Strategies and Limitations} \label{overload-strategy}Participants employed three strategies to cope with the overwhelming video presentation and information: (1) leveraging multimodal signifiers, (2) separating competing sources via pausing and rewinding, and (3) finding an attention anchor. We elaborate on each strategy and its limitations.


\textbf{\textit{Leveraging Multimodal Signifiers.}} Participants (P1, P4, P6-8, P10, P12-16) strategically directed their attention by following diverse visual (e.g., animations, bold text, color changes, instructor gestures) and auditory signifiers (e.g., key phrases like \textit{``the most important thing,''} shifts in tone and cadence, audience laughter). These signifiers helped them align speech with visual aid (\textit{``[The text] fades in when he's talking, so it helps you realize this is what he's talking about'' (P4)}), and filter information worth attending to (\textit{``I'm only paying attention to this because he said it's the most important thing'' (P11)).} The gaze data further validated this behavior, with participants' gaze following newly revealed content on slides or instructors' pointing gestures (Figure \ref{fig:gaze-challenges}f). 

\textit{\underline{Limitations.}} However, not all signifiers were equally effective: image signifiers (i.e., new images appearing on slides) attracted participants' gaze in 90.0\% of instances, compared to 41.3\% for new text and 40.1\% for instructor pointing gestures, highlighting the importance of signifier saliency in directing attention. As P11 noted: \textit{``When cool pictures appear I look at them. Text? Not so much.''} The effectiveness of signifiers also depended heavily on the video design and recording quality. Participants (P7-9, P12-13, P16) expressed frustrations over the lack of signifiers in videos, and two (P8, P13) noted that signifiers were easily lost when surrounding visuals were already overwhelming: \textit{``I didn't even notice [the bullet points] are being animated'' (P13).} Furthermore, poorly executed signifiers can bring additional burden to the participants (P1-2, P7, P11-12) by misdirecting their attention: \textit{``I've never seen anyone who has [used pointers] correctly... It's always random and distracting'' (P1).}

Additionally, P9 highlighted how lectures recorded in real-world classroom often failed to capture the signifiers designed for the in-person audiences: \textit{``If [the instructor] is pointing at anything on the screen, I can't see what they're pointing at... That is not recorded.''} 


\textbf{\textit{Separating Competing Information Sources.}} We found that participants (P2, P4, P7-8, P10, P12-13) also tried to alleviate cognitive overload by separating competing sources of information via pausing and replaying. For example, P2 paused the video (V8) to read the data in the table after listening to the speaker's explanation: \textit{``It took me a second to look at them all, so I had to pause after he was done talking... It's better for me to focus.''} Similarly, P10 paused the video to take notes on the slides before moving on to focusing on the speech, which served only as a supplement for filling any gap in knowledge. In total, participants paused 113 times ($Mean = 5.95$, $SD = 9.27$) and rewound 69 times ($Mean = 3.53$, $SD = 5.20$) across the video learning sessions, with 72.6\% of pauses concentrated on V2, V6, and V8---the three videos participants identified as exhibiting all three types of overwhelming presentation elements described in Section \ref{overwhelming-presentaiton}.

\underline{\textit{Limitations.}} However, separating different sources of information can make video watching both time-consuming and cognitively demanding. Four participants (P2, P8, P12-13) highlighted the additional effort required to watch such videos. For example, P2 spent a total of 10.7 minutes pausing and 4.2 minutes watching rewound content for the business video (V8), which almost doubled her video watching time from the original 16.5 minutes. In addition, two participants (P8, P13) mentioned the difficulty refocusing on video content after pausing: \textit{``Every time I have to pause it, it's hard to make it start again'' (P13).}  


\textbf{\textit{Finding an Attention Anchor.}} Seven participants (P1-2, P6, P8, P11-13) described a strategy of selectively focusing on a single element in the video when faced with visually overwhelming content, thereby reducing the cognitive load associated with processing ineffective visual aids. For example, P12 focused on the speaker when the slides were too complex. On a slide he described as having \textit{``too much text,''} 50.6\% of his fixation time was on the speaker, as shown in Figure \ref{fig:gaze-challenges}g. As he explained: \textit{``If the slide is full of stuff and I don’t want to look at it, I’m gonna focus on the speaker so that I don’t bury myself in the slide and lose my attention.''}


\underline{\textit{Limitations.}} The availability of an attention anchor depended on the video design. For example, P14 expressed his frustration when watching V7, a video lecture in the form of screen-recording of slides with voice-over: \textit{``I wanted to have that little area to look at... In the end, I had to look at a blank space, because I wanted to listen to what [the speaker] was saying, but it was hard to do that with all this stuff around'' (P14).} 



\subsection{Boredom from Understimulating Presentation}\label{understimulation} In contrast to cognitive overload, another major challenge highlighted by learners with ADHD lies at the opposite end of the spectrum---understimulation. Eleven participants (P1, P3-4, P6-9, P11-12, P14-15) described experiencing boredom across six videos (V1, V4-8). In the following, we unpack the video design factors that contribute to this sense of boredom and understimulation, and examine the coping strategies participants employed along with their limitations.

\subsubsection{Understimulating Lecture Designs.} \label{understimulating-designs} We found that understimulation during video learning could arise from both the speech and visual channels.

\textbf{\textit{Slow, Monotone Slide-Reading.}} Echoing prior work \cite{zhu2025character, adhdvideoaccess}, participants (P1, P3-5, P7-9, P11-12, P15) described monotone and slow-paced speech as a primary source of understimulation. In the context of video lectures, eight participants (P3, P7-9, P11-14) identified slide-reading as a major cause of monotone delivery. Though some participants (P8, P10-11, P15) acknowledged the tight speech-visual synchronization with slide reading could lower the cognitive load of reconciling multiple information sources, they noted that speakers typically flatten the prosodic variation, which is important for viewers with ADHD to identify important content. As P3 observed: \textit{``When someone reads, they don't put fluctuation in their voice, and they don't emphasize things. So it might be less overwhelming, but it's not effective for absorption.''} This flattened delivery left participants bored and inclined to disengage. As P9 shared: \textit{``[Slide reading] is monumentally stupid... makes things incredibly boring... I might as well just find something more interesting to do.''}

\textbf{\textit{Overly Static or Missing Visual Aids.}} 
We found that understimulation could also arise when the visual channel offered little variation in presentation. Six participants (P3-4, P6, P8, P12, P14) described videos with limited visual changes and stimulation as dull. For example, P8 criticized the health video (V6): \textit{``I can't believe it's just a black background with white text... And then I realized there was no variation in how these slides were presented. I was like, I don't really know if I want to go on.''} While she showed higher tolerance for content relevant to her field of study, she still expressed a wish for better-designed slides: \textit{``[These] slides weren't treated properly, but they could be, and it would be helpful if they were.''} Furthermore, seven participants (P2, P4, P9-10, P13, P15-16) shared the challenge of maintaining focus when visual aids failed to continuously illustrate the speaker's point when they went off-on-a-tangent: \textit{``There was one point in the lecture where [the speaker] went on a minute-long tangent about something... It's really hard to keep up when there's no visual supplement'' (P16).} 

\begin{figure*}
    \centering
    \includegraphics[width=0.95\linewidth, alt={Composite figure showing four examples of gaze trajectories overlaid on lecture video frames, illustrating how participants with ADHD experienced understimulation or attention disruption. Gaze is represented by circles (fixations, with larger circles indicating longer duration) connected by lines (saccades), with color progressing from red (earlier) to yellow (later). (a) Staring: prolonged fixation on a single point, indicating disengagement; (b) Switching attention: gaze shifts briefly to secondary elements to regain focus; (c) Revisiting: gaze returns to previously viewed content, reflecting difficulty recalling information; (d) Disorientation: scattered gaze across multiple elements following a disruption in content or structure.}]{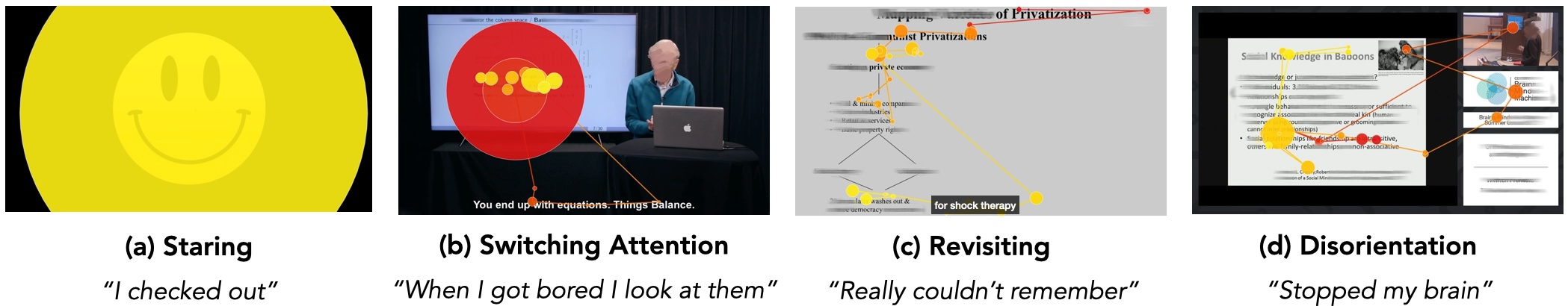}
    \caption{Examples of gaze trajectories when participants experienced \textbf{understimulation} or \textbf{attention disruption}.}
    \Description{Composite figure showing four examples of gaze trajectories overlaid on lecture video frames, illustrating how participants with ADHD experienced understimulation or attention disruption. Gaze is represented by circles (fixations, with larger circles indicating longer duration) connected by lines (saccades), with color progressing from red (earlier) to yellow (later). (a) Staring: prolonged fixation on a single point, indicating disengagement; (b) Switching attention: gaze shifts briefly to secondary elements to regain focus; (c) Revisiting: gaze returns to previously viewed content, reflecting difficulty recalling information; (d) Disorientation: scattered gaze across multiple elements following a disruption in content or structure.}
    \label{fig:gaze-challenges-understimulation}
\end{figure*}

Participants consistently described their experience during the understimulation as \textit{``checking out'' (P4)}---a state of reduced engagement that manifested in a unique gaze pattern. Seven participants (P1, P3-4, P6, P11-12, P14) exhibited this pattern, staring at a fixed point on the screen for a prolonged duration (\textit{Mean} = 3.26s, \textit{SD} = 1.78s), as shown in Figure \ref{fig:gaze-challenges-understimulation}a. 
These staring episodes were mainly reflected in two videos: 54.2\% in the art history video (V7), which had non-animated, singular visuals for most of its slides
, and 29.2\% in the mathematics video (V1), which was criticized as having monotone and slow speech as well as static visuals. 

\subsubsection{Coping Strategies and Limitations} \label{understimulation-strategy}Participants employed a variety of strategies to increase their stimulation in cases of boredom. We elaborate on their practices and their limitations below. 

\textbf{\textit{Attention Reboot through Secondary Stimulation.}} We found 10 participants (P1, P6-13, P15) voluntarily switched attention to other non-important visual components to increase stimulation. Unlike the attention anchor strategy (Section \ref{overload-strategy}), where participants narrowed gaze onto a single element, attention switching involved quick, temporary gaze movement onto secondary visual elements (e.g., miscellaneous information, caption) to reboot attention. 
As shown in Figure \ref{fig:gaze-challenges-understimulation}b, after P6 had zoned out (red circles), she quickly glanced at the captions to reboot attention, and went back to the slide with short, focal fixations (yellow circles). P1 added: \textit{``I paid more attention to the video because of the [secondary visuals]... If nothing's there, I would probably just look outside the window... [Switching my attention] gave me an opportunity to reset.''}  

\underline{\textit{Limitations.}} Participants' ability to switch attention within a video depended highly on the visual design of the video: overly simplistic visuals afforded nothing to switch to (especially when viewers disliked captions), while overly complex visuals might lead to distraction. For example, P3 illustrated the problem with overly simplistic visuals on a slide in the art history video (V7) containing only a smiley face: he stared at the smiley face for 6.7 seconds---the longest fixation across all participants' gaze data, indicating mind wandering \cite{negi2020fixation}: \textit{``There's nothing to decode or understand... I checked out to that degree where I didn't even realize where I was looking at... It would have been so much better if there's something else for me to pay attention to.''} Importantly, P1 highlighted that visual components for attention switching should be \textit{``familiar"} and \textit{``repetitive''} to reduce additional cognitive load: \textit{``I have to see something that I've seen before... I know what's gonna happen. I don't need to keep my attention on it.''}


\textbf{\textit{Intentional Mind Wandering.}} Seven participants (P3, P5, P7, P10, P13-14, P16) described intentionally reducing engagement on the video during segments they found unimportant or boring, while maintaining enough peripheral attention to return to more important content. For example, P7 described her mind wandering process when \textit{``entertain[ing]''} herself during a boring segment: \textit{``I'm looking at the man [in the video], like what's he doing? And then I wanted to know if my mom's friend's daughter and her boyfriend were still together, and I was thinking about how I wish I went to [a school].''} Despite appearing disengaged, participants reported that this was a regulated behavior: \textit{``Even when I'm zoning out, I pay enough attention where most of the time if [the speaker] is suddenly using very different words, my brain will just kind of refocus'' (P3).} P7 described this as a deliberate skill for allocating her limited attention: \textit{``You're not going to be interested in everything that you learn. So I've learned to be good at filtering stuff out.''} 

\underline{\textit{Limitations.}} While participants mentioned they were able to capture speech signals to re-attract their attention, the effectiveness of this strategy can be undermined by monotone speech---a primary cause for viewers to mind wander in the first place. As P14 shared: \textit{``If they're not a good presenter, then they are going to have this very stiff and monotone video... It's hard for me to go back because I won't be able to tell if something has changed.''}


\textbf{\textit{Multitasking.}} All participants mentioned multitasking as an important strategy to increase stimulation during video learning. Though our study only directly observed note-taking (P2, P5-7, P10, P12-16), participants described a broader set of multitasking activities they commonly employed: hands-on activities (e.g., coloring, crafting; P2-5, P7-8, P10, P13-14), online games (e.g., Solitaire, 2048; P1, P6), and light physical activity (P4, P9).

\underline{\textit{Limitations.}} Seven participants (P5-10, P16) acknowledged that multitasking might distract them from the learning task. For example, P10 highlighted his challenge with distinguishing important content for note-taking, which could cause her to overfocus on taking notes rather than understanding the content: \textit{``Most of my effort is being put to getting the information on the page rather than actively paying attention.''} In contrast, P8 shared the difficulty with note taking during hands-busy scenarios: \textit{``I really should have written that down... I just don't do it like I should.''} 


\textbf{\textit{Video Pace Adjustment.}} 
Echoing prior work \cite{adhdvideoaccess}, participants increased stimulation by speeding up the video (P2, P4-6, P9-11, P13, P15-16). These adjustments concentrated on specific understimulating videos that produced the most \textit{``checking out''} behaviors: the mathematics video (V1) was the only video that \textit{all} participants sped up, and was one of the only two videos (V1, V7) that had participants sped up to 2$\times$.  

\underline{\textit{Limitations.}} Faster playback speed could make speech feel unnatural (P2, P9) or pressure viewers to \textit{``rush for completion'' (P10)} 
rather than absorbing content (P6, P8, P10, P13). 

\subsection{Disrupted Attention from Unwanted Changes}
\label{disrupted-attention}
Eight participants (P1-2, P4, P10-11, P13, P15-16) reported that abrupt 
changes in videos could fragment their attention and disrupt comprehension. These disruptions arose from both visual transitions (e.g., scene switches) and content-level shifts (e.g., topic changes). 

\textbf{\textit{Forced Scene Switches.}} Five participants (P1-2, P4, P10, P15) described challenges with forced scene switches between different views (e.g., slides and speaker) for four videos (V1, V3-4, V6), making it difficult for them to remember content across switches given ADHD-related working memory challenges \cite{jacobson2011working}.  For example, P1 highlighted the trouble recalling the content on the slide after the video repeatedly switched between the slide and the speaker: \textit{``I don't like when he's going back and forth because I really couldn't remember what the slide was about.''} P1's gaze reflected this disruption when the political science video (V4) returned to the slide-view from the speaker-view: rather than picking up where she had left off (i.e., text at the bottom), she had to re-scan content she had previously viewed (Figure \ref{fig:gaze-challenges-understimulation}c). 
Similarly, P15 emphasized the feeling of disruption caused by such switches: \textit{``I really don't like switching back and forth... It disrupts my train of thought.''}

However, participant preferences for such 
scene switches in videos varied. 
Three participants (P5-6, P14) appreciated \textit{``more changes''}, which made the video more \textit{``dynamic''} and \textit{``interesting to look'' (P6).} Three others (P8, P9, P12) appreciated such changes when they functioned as signifiers for shifting attention between information sources. As P12 explained: \textit{``Switching [to the speaker] signals to me that I should be listening to what he's saying now... Versus when it zooms back to the slides, then there's something relevant on the slide, and I need to read through that.''} 

\textbf{\textit{Unexpected Topic Changes.}} Beyond visual transitions, participants (P2, P13) also experienced disruptions when speakers introduced abrupt or poorly signaled topic shifts. P13 highlighted the difficulty reconciling speech and visuals after a mid-sentence topic change in the neuroscience video (V2): \textit{``He had started saying a sentence and then branched away from it... That actually stopped my brain. I couldn't follow what he was saying, but I also couldn't continue reading the sentence to make sense of it.''} Reflecting this disorientation, her gaze scattered across multiple visual components immediately after the topic shift (Figure \ref{fig:gaze-challenges-understimulation}d). 
P2 described another disruption in the business video (V8): after thoroughly walking through a flowchart of the modern payment system, the speaker introduced a separate concept of digital wallet that was not reflected in the original flowchart. 
This suddenly appeared concept disrupted her flow of learning: \textit{``It threw me off because I wasn't expecting it.''} 


\textbf{\textit{Coping Strategies and Limitations.}} No effective strategies can overcome these abrupt content changes except for pausing and re-examining the content after disruption (P2, P4, P13). However, as highlighted in Section \ref{overload-strategy}, frequent pausing and restarting can also interrupt the video learning process, compounding rather than resolving the disruption.


\subsection{Confusion from Missing Key Content}
\label{confusion}
Nine participants (P6-9, P11-14, P16) experienced confusion stemming from the lack of key information, such as background knowledge, context, and learning goals. These issues affect learners broadly, but can be particularly challenging for individuals with ADHD, who may have greater difficulty organizing fragmented information into coherent mental representations \cite{kofler2018working}. 

\textbf{\textit{Insufficient Conceptual Context.}}
Nine participants (P6-9, P11-14, P16) described difficulties when four videos (V1-2, V5-6) introduced concepts without clearly defining key terms (e.g., acronyms) or situating them within a broader framework. For example, P8 complained about confusing terms used on a slide in V6: \textit{``I got so confused. It says genetic and environmental causes are each of two types. Two types of what? I didn't have any of that context... and then it just goes into examples, but what does that mean? ...It was a really tough slide for me.''} As a result, she paused the video to examine the whole slide, but still didn't manage to \textit{``get it.''} 

\textbf{\textit{Missing Goals and Expectations.}} Beyond missing definitions, three participants (P8, P11, P13) described an absence of clear learning goals. Without clarification on what information was important or how it fit into the broader topic, participants felt uncertain about how to engage with the content. As P13 noted: \textit{``I wasn't very primed for what I'm supposed to know.''} She highlighted this as a common challenge beyond our study, \textit{``It's like in a lot of university classes here, where the professor will just be talking, and I don't really know what I'm supposed to get out of what they're saying.''}

\textbf{\textit{Coping Strategies and Limitations.}} To fill the knowledge gap, participants reported turning to external resources (e.g., Google search) for clarification (P4, P9, P12, P15). However, leaving the video interface to seek external help introduced distraction: \textit{``If I go look something up on a web browser, then there's the possibility for me to open YouTube, open Amazon, or whatever'' (P9).} Moreover, external search could instead lead to information overload: \textit{``I feel like I end up down a rabbit hole, over-clarifying things that I don't need to because nobody's there to ask about it'' (P12).} 


\subsection{Addressing Higher-Level Video Learning Needs with Generative AI}
\label{ai-practices}
Beyond the sensory-level challenges unpacked in Sections \ref{cognitive-overload} to \ref{confusion}, participants surfaced higher-level learning needs that adaptations to individual video elements cannot fully resolve, including re-structuring inaccessible content, checking content understanding post-learning, and re-purposing learning with personal goals. Five participants (P3, P9, P12, P15, P16) addressed those needs via generative AI tools, yet more participants voiced concerns that made them hesitant to adopt AI. We elaborate on their practices, and highlight concerns and missed opportunities for AI-assisted video learning in the ADHD context.

\subsubsection{Addressing Learning Needs via Generative AI} We introduce participants’ use of generative AI tools to support their higher-level video learning needs.

\textbf{\textit{Re-structuring Inaccessible Content into Digestible Formats.}} Given the ADHD-specific challenge of organizing abstract and unstructured content \cite{kofler2018working}, all five participants who used AI in video learning relied on it to restructure content into more digestible formats such as summaries, highlights, and notes. While P16 used platform-embedded AI summarization (e.g., LinkedIn Learning), others actively prompted AI chatbots. For example, P3 described using an LLM to structure content for a video that he found \textit{``very obscure''} and \textit{``all over the place''}: \textit{``I gave the transcript to an LLM, and asked it to generate detailed notes so that I know how to structure my own.''} P12 also used AI to generate readings for preview, effectively transforming video learning into a format that better aligned with his preferences: \textit{``I just recreated my textbook.''}


\textbf{\textit{Checking Understanding Gaps through Post-learning Comparison.}} Two participants (P3, P12) used AI as a post-learning validator to check their understanding of the video. As P3 described: \textit{``I would take a picture of my notes, and I would ask [AI] to compare it with the transcript and see if there's anything missing.''} For P3, this cross-check served as a safeguard against the content he might have missed while checking out---a recurring concern for learners with ADHD as highlighted in Section \ref{understimulation}.

\textbf{\textit{Re-purposing Learning Based on Personal Goals.}} We also found one participant (P9) who used AI to re-purpose video learning into exam-driven learning: \textit{``Put [the learning objectives] into ChatGPT and say, make me a study guide... [ChatGPT] just knows to give me the buzzwords [for the exam].''} In this way, he was able to bypass the \textit{``very tedious''} video learning process and focus on exam preparation.


\subsubsection{Concerns: Authenticity, Accuracy, and the ADHD Burden of Validating Generated Content.} Despite the convenience that generative AI tools could afford, nine participants (P2-3, P5, P8-10, P13-14, P16) were hesitant to use AI in their learning due to two major concerns. The first was \textit{authenticity}: five participants (P2, P8-10, P13) worried that AI-generated summaries might not capture the instructor's intended educational goals and miss important information: \textit{``[AI] can summarize [the video], but then is that actually what I'm supposed to be pulling out of it?'' (P10)}. The second was \textit{accuracy}, with seven participants (P3, P5, P8-10, P14, P16) being skeptical of AI on domain-specific knowledge. These concerns echoed the challenges with AI-assisted learning faced by broader learners, including misalignment with instructor intent \cite{kasneci2023chatgpt} and inaccuracy on domain-specific content \cite{reihanian2024review}.

Despite similar concerns faced by general learners, individuals with ADHD encountered additional difficulties addressing such concerns. Although our participants (P3, P9, P12, P15) handled these concerns by cross-checking AI outputs against personal understanding (P3, P12), authoritative materials (P9, P12), and external search results (P12, P15), such triangulation processes were \textit{``tedious'' (P9)} and particularly difficult for individuals with ADHD to sustain. As P9 shared: \textit{``You're talking to ADHD people. I don't like to sit down and do boring, tedious, non-stimulating work... So once [the AI tools] have been right four or five times, you start to trust them on other stuff.''} He then highlighted the risks associated with such trust: \textit{``What if they are wrong the next time and I didn't bother to check?''}

\subsubsection{A Missed Opportunity: Externalizing Attention through AI-powered Feedback.} Interestingly, three participants (P8, P12, P14) expressed enthusiasm about the eye-tracking-based reflection method in our study, envisioning its use as a tool for monitoring and regulating their own attention. P12 found reviewing his gaze trajectories fun and helpful for optimizing his learning setup: \textit{``I can do my own study of what I pay attention to the best in what situations.''} Similarly, P14 shared the benefit of monitoring his attention: \textit{``I wish I could use this eye tracking in my personal life... Before, I was just a passenger to those distractions, but being aware of them can [help me] make the choice to focus.''} He envisioned a post-watching report, where AI summarizes engagement from behavioral data and highlight \textit{``parts I need to brush up.''} These insights suggest that externalizing attention could empower learners with ADHD to develop greater self-awareness and agency over their learning.








\section{Discussion}
This paper explores the experiences of viewers with ADHD during video learning and investigates how their experiences manifest in observable behaviors to inform the design of adaptive video learning systems. Through a retrospective think-aloud study with 16 participants with ADHD, we surfaced the video lecture design elements that hindered learning (e.g., overly dense visuals) and characterized how these challenges were reflected in viewer behaviors, such as misaligned gaze and prolonged fixations. We then unpacked the coping strategies that participants employed to navigate these challenges, both during video watching (e.g., attention reboot on secondary visuals when understimulated) and beyond (e.g., leveraging AI tools to restructure inaccessible content), while revealing the limitations of these strategies. 

In this section, we discuss the impact of our study in understanding and designing personalized and adaptive video learning in the context of ADHD with concrete design implications.

\subsection{Navigating Tensions in Video Adaptation Towards ADHD-friendly Video Learning}

Our findings surface two fundamental tensions that adaptive video learning systems for ADHD viewers need to navigate.

First, we identify a tension between \textit{efficiency} and \textit{fidelity} in supporting ADHD-friendly video learning, reflected in two distinct approaches: \textit{video-level adaptation} and \textit{beyond-video abstraction}. While beyond-video abstraction (e.g., AI-generated summaries) can substantially reduce cognitive load and improve learning efficiency, this approach risks missing important content and misaligning with the educator’s intent \cite{kasneci2023chatgpt}. In contrast, video-level adaptation preserves the original structure and grounding of the content, affording greater learner agency and trust, but could demand more effort and sustained attention. Our findings revealed that viewer agency and sense of control over their learning materials were critical in the context of ADHD video learning: even for participants who used AI-generated summaries, they still wanted to watch the video \textit{``to make sure [they] understand'' (P12).} This emphasis on agency echoes the broader self-regulated learning framework \cite{peel2019fundamentals,  schunk2005self}, which positioned learner control as central to effective learning. While prior work \cite{das2025towards} supporting ADHD video learning has focused on beyond-video abstraction, we encourage future research to also explore video-level adaptation as a method to support viewer agency and trust in the learning process. 

Another critical tension emerges between \textit{stimulation} and \textit{distraction} in \textit{video-level adaptation}. Stimulation-seeking has been characterized as a common behavioral trait of ADHD, with individuals often pursuing novel or intense stimuli to maintain arousal and engagement \cite{antrop2000stimulation, geissler2014hyperactivity}. However, individuals with ADHD can also be particularly vulnerable to distraction from irrelevant stimuli \cite{cassuto2013using, schneidt2018distraction}. Our findings extend this literature to the context of video adaptation, revealing that decisions about what to simplify, amplify, or remove from a video are shaped by a fundamental tension between stimulation and distraction, where both overly simple and complex designs can undermine engagement.  Simple presentations, while reducing distraction, often led to understimulation and \textit{``checking out,''} whereas visually rich designs could result in distraction and cognitive overload. 
Additionally, this tension cannot be resolved by a one-time design choice: participants' stimulation needs shifted dynamically with content difficulty, fatigue, and learning goals. While prior work has highlighted the importance of simple presentation designs for viewers with ADHD \cite{mcknight2010designing}, our findings suggest that effective video adaptation systems should not statically optimize for either simplicity or richness, but instead support \textit{adaptive modulation of stimulation} based on viewers' behavioral signals. 
We expand on the design implications in Section \ref{design-implication-system}.


\subsection{Informing ADHD Attention and Engagement with Gaze Behaviors}
Gaze data has been widely used in prior work as an indicator of attention and information processing \cite{borys2017eye, wang2021multi}, and our findings both align with and complicate these interpretations in the context of ADHD. Consistent with prior literature, we observed that certain gaze patterns corresponded to reduced engagement. For example, prolonged, stationary gaze often co-occurred with participants’ reports of mind wandering (Section \ref{understimulating-designs}), echoing prior work on longer fixation durations and reduced saccadic activity during attentional lapses \cite{negi2020fixation}. 

However, our findings also reveal the complexity and ambiguity of ADHD behavioral signals, where the same observable signal can indicate opposite internal states depending on the viewer's strategy and the surrounding content. On the one hand, we found that ostensibly attentive gaze patterns might not always reflect meaningful engagement, such as during mindless reading (Section \ref{overwhelming-presentaiton}), extending this finding from
prior literature \cite{reichle2010eye} with evidence from individuals with ADHD. On the other hand, we also found that gaze patterns commonly treated as markers of disengagement could reflect participants' intentional efforts to manage their attention. For example, while prior literature has highlighted dispersed gaze as an indicator of disengagement \cite{moiroud2025gaze, krasich2020eyes}, we observed that dispersed gaze patterns could also reflect participants' intentional effort to reset their focus via attention switching. Stationary gaze with prolonged fixations, despite indicating mind-wandering, could also be a strategic choice for individuals with ADHD to concentrate limited attention on more critical content (Section \ref{understimulation-strategy}). This finding underscores the complexity of ADHD gaze behaviors, and highlights that ADHD behavioral signals should not be interpreted in isolation when driving adaptation. We encourage future research to more deeply investigate the relationship between exhibited behaviors and level of engagement for individuals with ADHD, moving beyond binary classifications of attention and distraction toward a more nuanced, multi-signal and context-aware understanding of how individuals actively manage their engagement during complex tasks such as video learning.

\subsection{Design Implications for ADHD-friendly Video Learning} \label{design-implication} Our findings reveal multifaceted challenges that individuals with ADHD face during video learning. We recognize that addressing those challenges requires a collaborative effort between \textit{educators} and \textit{individualized adaptation systems}: educators can lower access barriers broadly through more inclusive design practices, while personalized adaptation systems address diverse individual needs that universal design cannot fully resolve. We discuss the design implications below. 

\subsubsection{For Educators: Toward ADHD-friendly Video Lecture Design.} We first explicate our recommendations for educators to make ADHD-friendly video lectures.

\textbf{\textit{Offering a Stable Attention Anchor.}} Our findings reveal that learners with ADHD frequently relied on a simple, consistent visual element to anchor their attention and focus on speech when slides became overwhelming. We recommend that educators ensure that a stable visual anchor, such as a persistent speaker view, is available throughout the lecture, and avoid formats that rely solely on slide screen-recordings with voice-over.

\textbf{\textit{Providing Multimodal Signifiers of Topic Shifts.}} Our findings highlight the importance of multimodal signifiers for learners with ADHD to follow dense materials and recollect their attention after distraction. We recommend educators explicitly mark topic transitions and key content using salient, multimodal signifiers, including verbal announcements (e.g., \textit{``most importantly''}) and visual changes (e.g., animations, text color changes). We highlight the importance of multimodality, as learners might be visually off-screen when managing their attention allocation (e.g., during multitasking, intentional mind-wandering). 

\textbf{\textit{Aligning Visual and Audio Content.}} We found that participants experienced a significant cognitive burden when prominent visuals (e.g., data visualizations) received insufficient verbal explanation. Educators should ensure that visually prominent elements are addressed in the speech and minimize the use of contextually irrelevant visuals that might draw learners' attention away from the core content. Additionally, when reading text on slides, educators should avoid branching away mid-sentence, as this forces learners to reconcile the remaining on-screen text with new speech content.

\textbf{\textit{Clarifying Context and Expectations.}} We found that participants experienced confusion when videos introduced concepts without sufficient context or clear learning goals, given their challenges with organizing fragmented information. We recommend educators open each lecture segment with a brief overview of learning objectives, and scaffold new concepts with sufficient background context or pointers to prerequisite materials (e.g., timestamped links for previous lectures). When applicable, we also recommend educators distinguish high-priority from supplementary content to support learners in mapping out key concepts and forming coherent mental representations of the material.

\textbf{\textit{Providing Balanced Visual Content Designs.}} Our findings highlight a core tension between understimulating and overwhelming visual designs for learners with ADHD. On a high level, we recommend educators design slides with an appropriate amount of text and a limited number of visuals, avoiding both text-heavy layouts and context-irrelevant imagery. This balance provides enough visual stimulation to sustain engagement while keeping individual slides focused and uncluttered. However, since the optimal balance could vary across individuals, we explain how adaptation systems can provide further support in Section \ref{design-implication-system}.

\subsubsection{For Adaptation System Designers: Toward Personalized Video Learning.}
\label{design-implication-system} 
While educators can make the effort to lower the access barriers for learners with ADHD, our findings also reveal that video learning challenges stem from dynamic mismatches between multimodal presentation and viewers’ \textit{individual} attentional needs. Importantly, participants demonstrated diverse (and sometimes conflicting) preferences for the video designs, highlighting the need to support personalized learning via video adaptation. Below, we outline design implications for future ADHD-friendly video adaptation systems.

\textbf{\textit{Enabling Behavior- and Context-aware Adaptation.}} Our findings revealed that viewers' attention states can be reflected in observable behaviors, but these signals can be complex and ambiguous (e.g., dispersed gaze can indicate distraction, confusion or intentional attention switch depending on contexts). This complexity suggests opportunities for adaptation systems to jointly consider behavioral signals and video context to infer viewers' engagement states and respond accordingly. For example, when dispersed gaze co-occurs with an overly static video segment, this might indicate a need to reboot attention via secondary stimulation, and the system could introduce lightweight visual variations (e.g., color changes in the text corresponding to current speech) to boost viewers' stimulation in a non-invasive way. Additionally, with multitasking being a critical coping strategy for ADHD video learning, adaptive systems could adjust presentation styles based on the viewers' multitasking activities. For example, the system could combine behavioral signals (e.g., front camera feed) and activity recognition models to detect viewer activity (e.g., hands-busy with crafting), and adapt the video presentation to support learning with partial attention (e.g., simplifying visual aid with only key information and allowing bookmarking via voice input).

\textbf{\textit{Offering Agency-preserving Video Designs.}} Our findings revealed the diverse preferences of individuals with ADHD towards video lecture designs. Future systems should therefore consider viewer preferences when presenting video content (e.g., by asking viewers to set default levels for elements such as text density, visual illustrations, and camera views). Beyond accommodating preferences, systems should also support viewers' agency to self-regulate attention. For example, as viewers actively regulate their attention through strategies such as rebooting with secondary visual components, adaptive systems should scaffold these strategies, such as adding user-selected secondary visuals that support attention resets during understimulating video segments.



\textbf{\textit{Designing Multimodal Signifiers to Guide Attention and Breaks.}} While educators could make the effort to design multimodal signifiers, learners with ADHD may still encounter poorly designed presentations without effective signifiers. This gap highlights an opportunity for future systems to augment or redesign signifiers to support viewers' attention. For example, the system can leverage vision-language models (VLMs) \cite{zhang2024vision} to identify slide regions that are semantically aligned with the current speech and dynamically highlight them (e.g., through bounding boxes or color changes), reducing the effort required to coordinate visual and auditory inputs. Furthermore, the system could combine learning objectives, slides, and transcript to classify the speech content into different topics and importance levels (e.g., key concepts vs. elaborations). Such classifications can further enable interaction techniques such as skipping off-tangent segments or providing lightweight cues (e.g., short beep) to recall attention when transitioning between topics.

\textbf{\textit{Supporting In-situ Clarification without Breaking Attention Flow.}} Our findings reveal that participants frequently experienced confusion during video watching, which introduced substantial risks of distraction when they left the video interface for clarification. This highlights the importance of integrating in-situ clarification mechanisms directly into the video interface. For example, future systems could infer confusing concepts via viewers' behaviors (e.g., hovering over a term for a long time), and offer clarifications from multiple sources (e.g., AI-generated vs. previous lecture timestamps) presented in a user-specified format (e.g., overlaid on video vs. at the side). 

\textbf{\textit{Incorporating Educator-aligned AI Support.}} 
The primary concern highlighted by participants regarding AI-powered video adaptation is the misalignment between the adapted learning content and the educator's intent. To minimize distortion and missing information (e.g., from simplified text), it is crucial to incorporate instructor's perspectives into both the adaptation process and outputs. For example, future systems could demonstrate bidirectional linking between adapted content and the original video, such as toggling back to original videos, anchoring summaries to video timestamps, and signaling omitted content on video timelines. Future systems could also explore methods to support educator-in-the-loop workflows, where educators can review or guide AI-generated adaptations (e.g., content presented with different text densities), ensuring that adaptations remain faithful to their teaching goals.

\subsection{Limitations and Future Work}

Our research has several limitations. First, our study was conducted in a lab environment. While this setup allowed us to capture real-time behavioral data and rich, timely reflections, it could also introduce observer effects \cite{baclawski2018observer}, where participants might alter their natural viewing behaviors due to being monitored. In addition, although we curated a diverse set of video lectures and took precautions to avoid assigning videos that participants explicitly disliked, the assigned video lectures may not fully reflect participants’ everyday learning contexts, where video lecture content is often self-selected and more personally relevant. Future work should thus explore longitudinal field deployments that integrate behavioral tracking systems into real-world learning environments to understand how ADHD-related challenges are reflected within authentic video learning contexts.

Second, our study primarily adopted an exploratory approach to qualitatively understand viewers' experiences with a relatively limited number of participants. While our approach offers in-depth understanding of viewers' learning experiences, it did not systematically control video variation and therefore cannot measure the quantitative effect of any individual design dimension. Additionally, our post-viewing quizzes were designed primarily to encourage serious engagement with the learning task. Although we found quiz performance to be positively correlated with participants' perceived content absorption, it cannot capture broader learning outcomes or long-term retention, which is a key indicator of effective learning \cite{ausubel2012acquisition}. Future work could conduct larger-scale controlled studies that systematically vary individual design dimensions (e.g., text density, instructor presence), while incorporating more explicit and scalable measures of attention and engagement \cite{weinstein2018mind, srivastava2019continuous}, as well as assessments of short-term learning outcomes and long-term retention \cite{abbasi2014measuring, ashby2004monitoring}, to quantify how different video designs affect learning for viewers with ADHD.

\begin{acks}
This work was partially supported by an Apple Seed Grant. 
\end{acks}

\bibliographystyle{ACM-Reference-Format}
\bibliography{main}

@String{Computing = "Computing" }

@String{Computer = "{IEEE} Computer" }

@String{Academic = "Academic Press" }

@String{Springer = "Springer-Verlag" }

@article{antrop2000stimulation,
  title={Stimulation seeking and hyperactivity in children with ADHD},
  author={Antrop, Inge and Roeyers, Herbert and Van Oost, Paulette and Buysse, Ann},
  journal={The Journal of Child Psychology and Psychiatry and Allied Disciplines},
  volume={41},
  number={2},
  pages={225--231},
  year={2000},
  publisher={Cambridge University Press}
}

@article{roberts2012constraints,
  title={Constraints on information processing capacity in adults with ADHD.},
  author={Roberts, Walter and Milich, Richard and Fillmore, Mark T},
  journal={Neuropsychology},
  volume={26},
  number={6},
  pages={695},
  year={2012},
  publisher={American Psychological Association}
}

@inproceedings{adhdvideoaccess,
author = {Jiang, Lucy and Ko, Woojin and Yuan, Shirley and Shende, Tanisha and Azenkot, Shiri},
title = {Shifting the Focus: Exploring Video Accessibility Strategies and Challenges for People with ADHD},
year = {2025},
isbn = {9798400713941},
publisher = {Association for Computing Machinery},
address = {New York, NY, USA},
url = {https://doi.org/10.1145/3706598.3713637},
doi = {10.1145/3706598.3713637},
booktitle = {Proceedings of the 2025 CHI Conference on Human Factors in Computing Systems},
articleno = {561},
numpages = {16},
location = {
},
series = {CHI '25}
}

@article{groen2020testing,
  title={Testing the relation between ADHD and hyperfocus experiences},
  author={Groen, Yvonne and Priegnitz, Ulrike and Fuermaier, Anselm BM and Tucha, Lara and Tucha, Oliver and Aschenbrenner, Steffen and Weisbrod, Matthias and Pimenta, Miguel Garcia},
  journal={Research in Developmental Disabilities},
  volume={107},
  pages={103789},
  year={2020},
  publisher={Elsevier}
}

@article{braun2006using,
  title={Using thematic analysis in psychology},
  author={Braun, Virginia and Clarke, Victoria},
  journal={Qualitative research in psychology},
  volume={3},
  number={2},
  pages={77--101},
  year={2006},
  publisher={Taylor \& Francis}
}

@misc{React,
  author       = {{React Team}},
  title        = {React},
  year         = {2025},
  howpublished = {\url{https://react.dev/}},
  note         = {Accessed: 2025-07-17}
}

@article{chorianopoulos2018taxonomy,
  title={A taxonomy of asynchronous instructional video styles},
  author={Chorianopoulos, Konstantinos},
  journal={International Review of Research in Open and Distributed Learning},
  volume={19},
  number={1},
  year={2018},
  publisher={{\'E}rudit}
}

@article{doernberg2016neurodevelopmental,
  title={Neurodevelopmental disorders (asd and adhd): Dsm-5, icd-10, and icd-11},
  author={Doernberg, Ellen and Hollander, Eric},
  journal={CNS spectrums},
  volume={21},
  number={4},
  pages={295--299},
  year={2016},
  publisher={Cambridge University Press}
}

@article{schneidt2018distraction,
  title={Distraction by salient stimuli in adults with attention-deficit/hyperactivity disorder: Evidence for the role of task difficulty in bottom-up and top-down processing},
  author={Schneidt, Alexander and Jusyte, Aiste and Rauss, Karsten and Sch{\"o}nenberg, Michael},
  journal={Cortex},
  volume={101},
  pages={206--220},
  year={2018},
  publisher={Elsevier}
}

@article{loe2007academic,
  title={Academic and educational outcomes of children with ADHD},
  author={Loe, Irene M and Feldman, Heidi M},
  journal={Journal of pediatric psychology},
  volume={32},
  number={6},
  pages={643--654},
  year={2007},
  publisher={Society of Pediatric Psychology}
}

@article{nadeau2005career,
  title={Career choices and workplace challenges for individuals with ADHD},
  author={Nadeau, Kathleen G},
  journal={Journal of Clinical Psychology},
  volume={61},
  number={5},
  pages={549--563},
  year={2005},
  publisher={Wiley Online Library}
}

@article{arnold2020long,
  title={Long-term outcomes of ADHD: academic achievement and performance},
  author={Arnold, L Eugene and Hodgkins, Paul and Kahle, Jennifer and Madhoo, Manisha and Kewley, Geoff},
  journal={Journal of attention disorders},
  volume={24},
  number={1},
  pages={73--85},
  year={2020},
  publisher={Sage Publications Sage CA: Los Angeles, CA}
}

@article{jacobson2011working,
  title={Working memory influences processing speed and reading fluency in ADHD},
  author={Jacobson, Lisa A and Ryan, Matthew and Martin, Rebecca B and Ewen, Joshua and Mostofsky, Stewart H and Denckla, Martha B and Mahone, E Mark},
  journal={Child neuropsychology},
  volume={17},
  number={3},
  pages={209--224},
  year={2011},
  publisher={Taylor \& Francis}
}

@article{eagle2024something,
  title={“It was something I naturally found worked and heard about later”: An Investigation of Body Doubling with Neurodivergent Participants},
  author={Eagle, Tessa and Baltaxe-Admony, Leya Breanna and Ringland, Kathryn E},
  journal={ACM Transactions on Accessible Computing},
  volume={17},
  number={3},
  pages={1--30},
  year={2024},
  publisher={ACM New York, NY}
}

@article{rucklidge2002neuropsychological,
  title={Neuropsychological profiles of adolescents with ADHD: Effects of reading difficulties and gender},
  author={Rucklidge, Julia J and Tannock, Rosemary},
  journal={Journal of child psychology and psychiatry},
  volume={43},
  number={8},
  pages={988--1003},
  year={2002},
  publisher={Wiley Online Library}
}

@article{kim2014visual,
  title={Visual function and color vision in adults with Attention-Deficit/Hyperactivity Disorder},
  author={Kim, Soyeon and Chen, Samantha and Tannock, Rosemary},
  journal={Journal of optometry},
  volume={7},
  number={1},
  pages={22--36},
  year={2014},
  publisher={Elsevier}
}

@article{levenberg2023learning,
  title={Learning from recorded lectures: Perceptions of students with ADHD},
  author={Levenberg, Ariella and Reesh, Suzan Abu},
  journal={Journal of attention disorders},
  volume={27},
  number={9},
  pages={960--972},
  year={2023},
  publisher={SAGE Publications Sage CA: Los Angeles, CA}
}

@article{blomberg2021effects,
  title={The effects of working memory load on auditory distraction in adults with attention deficit hyperactivity disorder},
  author={Blomberg, Rina and Johansson Capusan, Andrea and Signoret, Carine and Danielsson, Henrik and R{\"o}nnberg, Jerker},
  journal={Frontiers in Human Neuroscience},
  volume={15},
  pages={771711},
  year={2021},
  publisher={Frontiers Media SA}
}

@article{cassuto2013using,
  title={Using environmental distractors in the diagnosis of ADHD},
  author={Cassuto, Hanoch and Ben-Simon, Anat and Berger, Itai},
  journal={Frontiers in human neuroscience},
  volume={7},
  pages={805},
  year={2013},
  publisher={Frontiers Media SA}
}

@inproceedings{Lalwani2025,
author = {Lalwani, Himanshi and Saleh, Mira and Salam, Hanan},
title = {A Study Companion for Productivity: Exploring the Role of a Social Robot for College Students with ADHD},
year = {2025},
publisher = {IEEE Press},
booktitle = {Proceedings of the 2025 ACM/IEEE International Conference on Human-Robot Interaction},
pages = {1438–1442},
numpages = {5},
location = {Melbourne, Australia},
series = {HRI '25}
}

@inproceedings{cuber2024,
author = {Cuber, Isabelle and Goncalves De Souza, Juliana G and Jacobs, Irene and Lowman, Caroline and Shepherd, David and Fritz, Thomas and Langberg, Joshua M},
title = {Examining the Use of VR as a Study Aid for University Students with ADHD},
year = {2024},
isbn = {9798400703300},
publisher = {Association for Computing Machinery},
address = {New York, NY, USA},
url = {https://doi.org/10.1145/3613904.3643021},
doi = {10.1145/3613904.3643021},
booktitle = {Proceedings of the 2024 CHI Conference on Human Factors in Computing Systems},
articleno = {65},
numpages = {16},
location = {Honolulu, HI, USA},
series = {CHI '24}
}

@article{das2021,
author = {Das, Maitraye and Tang, John and Ringland, Kathryn E. and Piper, Anne Marie},
title = {Towards Accessible Remote Work: Understanding Work-from-Home Practices of Neurodivergent Professionals},
year = {2021},
issue_date = {April 2021},
publisher = {Association for Computing Machinery},
address = {New York, NY, USA},
volume = {5},
number = {CSCW1},
url = {https://doi.org/10.1145/3449282},
doi = {10.1145/3449282},
journal = {Proc. ACM Hum.-Comput. Interact.},
month = apr,
articleno = {183},
numpages = {30}
}

@inproceedings{mcknight2010designing,
  title={Designing for ADHD in search of guidelines},
  author={McKnight, Lorna},
  booktitle={IDC 2010 digital technologies and marginalized youth workshop},
  volume={30},
  year={2010}}

@article{tucha2017sustained,
  title={Sustained attention in adult ADHD: time-on-task effects of various measures of attention},
  author={Tucha, Lara and Fuermaier, Anselm BM and Koerts, Janneke and Buggenthin, Rieka and Aschenbrenner, Steffen and Weisbrod, Matthias and Thome, Johannes and Lange, Klaus W and Tucha, Oliver},
  journal={Journal of neural transmission},
  volume={124},
  pages={39--53},
  year={2017},
  publisher={Springer}
}

@ARTICLE{Wilens2010-hq,
  title     = "Understanding attention-deficit/hyperactivity disorder from
               childhood to adulthood",
  author    = "Wilens, Timothy E and Spencer, Thomas J",
  journal   = "Postgrad. Med.",
  publisher = "Informa UK Limited",
  volume    =  122,
  number    =  5,
  pages     = "97--109",
  month     =  sep,
  year      =  2010,
  language  = "en"
}

@inproceedings{10.1145/2858036.2858157,
author = {Sonne, Tobias and M\"{u}ller, J\"{o}rg and Marshall, Paul and Obel, Carsten and Gr\o{}nb\ae{}k, Kaj},
title = {Changing Family Practices with Assistive Technology: MOBERO Improves Morning and Bedtime Routines for Children with ADHD},
year = {2016},
isbn = {9781450333627},
publisher = {Association for Computing Machinery},
address = {New York, NY, USA},
url = {https://doi.org/10.1145/2858036.2858157},
doi = {10.1145/2858036.2858157},
booktitle = {Proceedings of the 2016 CHI Conference on Human Factors in Computing Systems},
pages = {152–164},
numpages = {13},
location = {San Jose, California, USA},
series = {CHI '16}
}

@article{Sobanski2006,
  title = {Psychiatric comorbidity in adults with attention-deficit/hyperactivity disorder (ADHD)},
  volume = {256},
  ISSN = {1433-8491},
  url = {http://dx.doi.org/10.1007/s00406-006-1004-4},
  DOI = {10.1007/s00406-006-1004-4},
  number = {S1},
  journal = {European Archives of Psychiatry and Clinical Neuroscience},
  publisher = {Springer Science and Business Media LLC},
  author = {Sobanski,  Esther},
  year = {2006},
  month = sep,
  pages = {i26–i31}
}

@article{reale2017,
  title = {Comorbidity prevalence and treatment outcome in children and adolescents with ADHD},
  volume = {26},
  ISSN = {1435-165X},
  url = {http://dx.doi.org/10.1007/s00787-017-1005-z},
  DOI = {10.1007/s00787-017-1005-z},
  number = {12},
  journal = {European Child \&amp; Adolescent Psychiatry},
  publisher = {Springer Science and Business Media LLC},
  author = {Reale,  Laura and Bartoli,  Beatrice and Cartabia,  Massimo and Zanetti,  Michele and Costantino,  Maria Antonella and Canevini,  Maria Paola and Termine,  Cristiano and Bonati,  Maurizio},
  year = {2017},
  month = may,
  pages = {1443–1457}
}

@article{Wittenberg2021,
  title = {The (minimal) persuasive advantage of political video over text},
  volume = {118},
  ISSN = {1091-6490},
  url = {http://dx.doi.org/10.1073/pnas.2114388118},
  DOI = {10.1073/pnas.2114388118},
  number = {47},
  journal = {Proceedings of the National Academy of Sciences},
  publisher = {Proceedings of the National Academy of Sciences},
  author = {Wittenberg,  Chloe and Tappin,  Ben M. and Berinsky,  Adam J. and Rand,  David G.},
  year = {2021},
  month = nov 
}

@article{Li2022,
  title = {Impact of information timeliness and richness on public engagement on social media during COVID-19 pandemic: An empirical investigation based on NLP and machine learning},
  volume = {162},
  ISSN = {0167-9236},
  url = {http://dx.doi.org/10.1016/j.dss.2022.113752},
  DOI = {10.1016/j.dss.2022.113752},
  journal = {Decision Support Systems},
  publisher = {Elsevier BV},
  author = {Li,  Kai and Zhou,  Cheng and Luo,  Xin (Robert) and Benitez,  Jose and Liao,  Qinyu},
  year = {2022},
  month = nov,
  pages = {113752}
}

@article{Salari2023,
  title = {The global prevalence of ADHD in children and adolescents: a systematic review and meta-analysis},
  volume = {49},
  ISSN = {1824-7288},
  url = {http://dx.doi.org/10.1186/s13052-023-01456-1},
  DOI = {10.1186/s13052-023-01456-1},
  number = {1},
  journal = {Italian Journal of Pediatrics},
  publisher = {Springer Science and Business Media LLC},
  author = {Salari,  Nader and Ghasemi,  Hooman and Abdoli,  Nasrin and Rahmani,  Adibeh and Shiri,  Mohammad Hossain and Hashemian,  Amir Hossein and Akbari,  Hakimeh and Mohammadi,  Masoud},
  year = {2023},
  month = apr 
}

@ARTICLE{Kokoc2020-if,
  title     = "Effects of sustained attention and video lecture types on
               learning performances",
  author    = "Koko{\c c}, Mehmet and IIgaz, Hale and Altun, Arif",
  journal   = "Educ. Technol. Res. Dev.",
  publisher = "Springer Science and Business Media LLC",
  volume    =  68,
  number    =  6,
  pages     = "3015--3039",
  month     =  dec,
  year      =  2020,
  language  = "en"
}

@inproceedings{zhu2025character,
author = {Zhu, Hanxiu ‘Hazel’ and Senthil Kumar, Avanthika and Zhao, Sihang and Wang, Ru and Tong, Xin and Zhao, Yuhang},
title = {Characterizing Collective Efforts in Content Sharing and Quality Control for ADHD-relevant Content on Video-sharing Platforms},
year = {2025},
isbn = {9798400706769},
publisher = {Association for Computing Machinery},
address = {New York, NY, USA},
url = {https://doi.org/10.1145/3663547.3746387},
doi = {10.1145/3663547.3746387},
booktitle = {Proceedings of the 27th International ACM SIGACCESS Conference on Computers and Accessibility},
articleno = {69},
numpages = {15},
location = {
},
series = {ASSETS '25}
}

@inproceedings{focusview,
author = {Zhu, Hanxiu ‘Hazel’ and Chen, Ruijia and Zhao, Yuhang},
title = {FocusView: Understanding and Customizing Informational Video Watching Experiences for Viewers with ADHD},
year = {2025},
isbn = {9798400706769},
publisher = {Association for Computing Machinery},
address = {New York, NY, USA},
url = {https://doi.org/10.1145/3663547.3746386},
doi = {10.1145/3663547.3746386},
booktitle = {Proceedings of the 27th International ACM SIGACCESS Conference on Computers and Accessibility},
articleno = {70},
numpages = {18},
location = {
},
series = {ASSETS '25}
}

@inproceedings{sonne2015,
author = {Sonne, Tobias and Obel, Carsten and Gr\o{}nb\ae{}k, Kaj},
title = {Designing Real Time Assistive Technologies: A Study of Children with ADHD},
year = {2015},
isbn = {9781450336734},
publisher = {Association for Computing Machinery},
address = {New York, NY, USA},
url = {https://doi.org/10.1145/2838739.2838815},
doi = {10.1145/2838739.2838815},
booktitle = {Proceedings of the Annual Meeting of the Australian Special Interest Group for Computer Human Interaction},
pages = {34–38},
numpages = {5},
location = {Parkville, VIC, Australia},
series = {OzCHI '15}
}

@article{batanero2014considering,
  title={Considering student personal needs and preferences and accessible learning objects to adapt Moodle learning platform},
  author={Batanero, Concha and Ot{\'o}n, Salvador and Alonso, Jaime and Holvikivi, Jaana},
  year={2014}
}

@article{Standen2020,
author = {Standen, Penelope J. and Brown, David J. and Taheri, Mohammad and Galvez Trigo, Maria J. and Boulton, Helen and Burton, Andrew and Hallewell, Madeline J. and Lathe, James G. and Shopland, Nicholas and Blanco Gonzalez, Maria A. and Kwiatkowska, Gosia M. and Milli, Elena and Cobello, Stefano and Mazzucato, Annaleda and Traversi, Marco and Hortal, Enrique},
title = {An evaluation of an adaptive learning system based on multimodal affect recognition for learners with intellectual disabilities},
journal = {British Journal of Educational Technology},
volume = {51},
number = {5},
pages = {1748-1765},
doi = {https://doi.org/10.1111/bjet.13010},
url = {https://bera-journals.onlinelibrary.wiley.com/doi/abs/10.1111/bjet.13010},
eprint = {https://bera-journals.onlinelibrary.wiley.com/doi/pdf/10.1111/bjet.13010},
year = {2020}
}

@ARTICLE{Vile_Junod2006-tk,
  title     = "Classroom observations of students with and without {ADHD}:
               Differences across types of engagement",
  author    = "Vile Junod, Rosemary E and DuPaul, George J and Jitendra, Asha K
               and Volpe, Robert J and Cleary, Kristi S",
  journal   = "J. Sch. Psychol.",
  publisher = "Elsevier BV",
  volume    =  44,
  number    =  2,
  pages     = "87--104",
  month     =  apr,
  year      =  2006,
  language  = "en"
}

@INPROCEEDINGS{Thawalampola2024,
  author={Thawalampola, Oshada and Jayasuriya, Danuja and Kariyawasam, Supun and Makawita, Madushi and Wijendra, Dinuka and Joseph, Jenny},
  booktitle={2024 6th International Conference on Advancements in Computing (ICAC)}, 
  title={Adaptive Learning Tool to Enhance Educational Outcomes for Students with Inattentive Attention Deficit Hyperactivity Disorder (ADHD)}, 
  year={2024},
  volume={},
  number={},
  pages={462-467},
  doi={10.1109/ICAC64487.2024.10850982}}

@inproceedings{yadav2025,
author = {Yadav, Saumya},
title = {Decoding Attention in Children with Attention Deficit Hyperactivity Disorder through Multimodal Analysis for Digital Learning},
year = {2025},
isbn = {9798400713958},
publisher = {Association for Computing Machinery},
address = {New York, NY, USA},
url = {https://doi.org/10.1145/3706599.3707607},
doi = {10.1145/3706599.3707607},
booktitle = {Proceedings of the Extended Abstracts of the CHI Conference on Human Factors in Computing Systems},
articleno = {828},
numpages = {6},
location = {
},
series = {CHI EA '25}
}

@article{emam2025enhancing,
  title={Enhancing Educational Videos for ADHD Learners: A Review of Multimedia Design and Deep Learning Frameworks},
  author={Emam, Alshefaa Mohamed and ElSayed, Eman and Gallab, Mai},
  journal={International Journal of Applied Intelligent Computing and Informatics},
  volume={1},
  number={2},
  pages={63--70},
  year={2025},
  publisher={Banha University, Faculty of Computers and Artificial Intelligence}
}

@article{alpert2019video,
  title={Video use in lecture classes: Current practices, student perceptions and preferences},
  author={Alpert, Frank and Hodkinson, Chris S},
  journal={Education+ Training},
  volume={61},
  number={1},
  pages={31--45},
  year={2019},
  publisher={Emerald Publishing Limited}
}

@article{reaser2007learning,
  title={The learning and study strategies of college students with ADHD},
  author={Reaser, Abigail and Prevatt, Frances and Petscher, Yaacov and Proctor, Briley},
  journal={Psychology in the Schools},
  volume={44},
  number={6},
  pages={627--638},
  year={2007},
  publisher={Wiley Online Library}
}

@article{noetel2021video,
  title={Video improves learning in higher education: A systematic review},
  author={Noetel, Michael and Griffith, Shantell and Delaney, Oscar and Sanders, Taren and Parker, Philip and del Pozo Cruz, Borja and Lonsdale, Chris},
  journal={Review of educational research},
  volume={91},
  number={2},
  pages={204--236},
  year={2021},
  publisher={Sage Publications Sage CA: Los Angeles, CA}
}

@article{long2023review,
  title={A review on the use of video in education: Advantages and disadvantages},
  author={Long, Ong Ace Hong Ong and Abd Halim, Noor Dayana and Hanid, Mohd Fadzil Abdul},
  journal={Innovative Teaching and Learning Journal},
  volume={7},
  number={2},
  pages={25--40},
  year={2023}
}

@article{mayer2024past,
  title={The past, present, and future of the cognitive theory of multimedia learning},
  author={Mayer, Richard E},
  journal={Educational Psychology Review},
  volume={36},
  number={1},
  pages={8},
  year={2024},
  publisher={Springer}
}

@article{cceken2022multimedia,
  title={Multimedia learning principles in different learning environments: A systematic review},
  author={{\c{C}}eken, Bur{\c{c}} and Ta{\c{s}}k{\i}n, Naz{\i}m},
  journal={Smart Learning Environments},
  volume={9},
  number={1},
  pages={19},
  year={2022},
  publisher={Springer}
}

@book{clark2023learning,
  title={E-learning and the science of instruction: Proven guidelines for consumers and designers of multimedia learning},
  author={Clark, Ruth C and Mayer, Richard E},
  year={2023},
  publisher={john Wiley \& sons}
}

@article{alshaikh2024implementation,
  title={The implementation of the cognitive theory of multimedia learning in the design and evaluation of an AI educational video assistant utilizing large language models},
  author={AlShaikh, Rana and Al-Malki, Norah and Almasre, Maida},
  journal={Heliyon},
  volume={10},
  number={3},
  year={2024},
  publisher={Elsevier}
}

@article{fyfield2022improving,
  title={Improving instructional video design: A systematic review},
  author={Fyfield, Matthew and Henderson, Michael and Phillips, Michael},
  journal={Australasian Journal of Educational Technology},
  volume={38},
  number={3},
  pages={155--183},
  year={2022}
}

@article{cavanagh2023using,
  title={Using commonly-available technologies to create online multimedia lessons through the application of the Cognitive Theory of Multimedia Learning},
  author={Cavanagh, Thomas M and Kiersch, Christa},
  journal={Educational technology research and development},
  volume={71},
  number={3},
  pages={1033--1053},
  year={2023},
  publisher={Springer}
}

@article{chen2024automatic,
  title={Automatic generation of multimedia teaching materials based on generative AI: Taking Tang poetry as an example},
  author={Chen, Xu and Wu, Di},
  journal={IEEE transactions on learning technologies},
  volume={17},
  pages={1327--1340},
  year={2024},
  publisher={IEEE}
}

@article{namestovski2022framework,
  title={Framework for preparation of engaging online educational materials—a cognitive approach},
  author={Namestovski, {\v{Z}}olt and Kovari, Attila},
  journal={Applied Sciences},
  volume={12},
  number={3},
  pages={1745},
  year={2022},
  publisher={MDPI}
}

@incollection{kirschner2023toward,
  title={Toward a cognitive theory of multimedia assessment (CTMMA)},
  author={Kirschner, Paul A and Park, Babette and Malone, Sarah and Jarodzka, Halszka},
  booktitle={Learning, design, and technology: An international compendium of theory, research, practice, and policy},
  pages={153--175},
  year={2023},
  publisher={Springer}
}

@article{mahak2025academic,
  title={Academic anxiety and cognitive reflection in neurodivergence based on evidence from a large international sample},
  author={Mahak, Sheeza and Malone, Stephanie and Elsherif, Mahmoud and Hand, Christopher J and Morsanyi, Kinga},
  journal={Scientific Reports},
  volume={15},
  number={1},
  pages={37522},
  year={2025},
  publisher={Nature Publishing Group UK London}
}

@article{mayes2020sluggish,
  title={Sluggish cognitive tempo in autism, ADHD, and neurotypical child samples},
  author={Mayes, Susan D and Calhoun, Susan L and Waschbusch, Daniel A},
  journal={Research in Autism Spectrum Disorders},
  volume={79},
  pages={101678},
  year={2020},
  publisher={Elsevier}
}

@article{hite2021describing,
  title={Describing the experiences of students with ADHD learning science content with emerging technologies},
  author={Hite, Rebecca and Childers, Gina and Jones, Gail and Corin, Elysa and Pereyra, Mariana},
  journal={Journal of Science Education for Students with Disabilities},
  volume={24},
  number={1},
  pages={12},
  year={2021}
}

@inproceedings{das2025towards,
  title={Towards a Technology that Improves Focus, Comprehension, and Retention in E-learning for Higher Education students with ADHD},
  author={Das, Aishee and Constantin, Aurora and Imperatore, Gennaro},
  booktitle={Proceedings of the 27th International ACM SIGACCESS Conference on Computers and Accessibility},
  pages={1--4},
  year={2025}
}

@article{martin2020systematic,
  title={Systematic review of adaptive learning research designs, context, strategies, and technologies from 2009 to 2018},
  author={Martin, Florence and Chen, Yan and Moore, Robert L and Westine, Carl D},
  journal={Educational Technology Research and Development},
  volume={68},
  number={4},
  pages={1903--1929},
  year={2020},
  publisher={Springer}
}

@article{halkiopoulos2024leveraging,
  title={Leveraging AI in e-learning: Personalized learning and adaptive assessment through cognitive neuropsychology—A systematic analysis},
  author={Halkiopoulos, Constantinos and Gkintoni, Evgenia},
  journal={Electronics},
  volume={13},
  number={18},
  pages={3762},
  year={2024},
  publisher={MDPI}
}

@article{van2003retrospective,
  title={Retrospective vs. concurrent think-aloud protocols: testing the usability of an online library catalogue},
  author={Van Den Haak, Maaike and De Jong, Menno and Jan Schellens, Peter},
  journal={Behaviour \& information technology},
  volume={22},
  number={5},
  pages={339--351},
  year={2003},
  publisher={Taylor \& Francis}
}

@inproceedings{alhadreti2018rethinking,
  title={Rethinking thinking aloud: A comparison of three think-aloud protocols},
  author={Alhadreti, Obead and Mayhew, Pam},
  booktitle={Proceedings of the 2018 CHI conference on human factors in computing systems},
  pages={1--12},
  year={2018}
}

@article{hoza2001academic,
  title={Academic task persistence of normally achieving ADHD and control boys: Self-evaluations, and attributions.},
  author={Hoza, Betsy and Waschbusch, Daniel A and Owens, Julie Sarno and Pelham, William E and Kipp, Heidi},
  journal={Journal of consulting and clinical psychology},
  volume={69},
  number={2},
  pages={271},
  year={2001},
  publisher={American Psychological Association}
}

@article{modesto2013motivation,
  title={Are motivation deficits underestimated in patients with ADHD? A review of the literature},
  author={Modesto-Lowe, Vania and Chaplin, Margaret and Soovajian, Victoria and Meyer, Andrea},
  journal={Postgraduate medicine},
  volume={125},
  number={4},
  pages={47--52},
  year={2013},
  publisher={Taylor \& Francis}
}

@article{kofler2018working,
  title={Working memory and organizational skills problems in ADHD},
  author={Kofler, Michael J and Sarver, Dustin E and Harmon, Sherelle L and Moltisanti, Allison and Aduen, Paula A and Soto, Elia F and Ferretti, Nicole},
  journal={Journal of child psychology and psychiatry},
  volume={59},
  number={1},
  pages={57--67},
  year={2018},
  publisher={Wiley Online Library}
}

@article{bikic2017meta,
  title={Meta-analysis of organizational skills interventions for children and adolescents with Attention-Deficit/Hyperactivity Disorder},
  author={Bikic, Aida and Reichow, Brian and McCauley, Spencer A and Ibrahim, Karim and Sukhodolsky, Denis G},
  journal={Clinical psychology review},
  volume={52},
  pages={108--123},
  year={2017},
  publisher={Elsevier}
}

@article{marchetta2008sustained,
  title={Sustained and focused attention deficits in adult ADHD},
  author={Marchetta, Natalie DJ and Hurks, Petra PM and De Sonneville, Leo MJ and Krabbendam, Lydia and Jolles, Jelle},
  journal={Journal of Attention Disorders},
  volume={11},
  number={6},
  pages={664--676},
  year={2008},
  publisher={Sage Publications Sage CA: Los Angeles, CA}
}

@article{karnad2013neurodiversity,
  title={Neurodiversity and lecture recordings},
  author={Karnad, Arun and Bond, Steve},
  year={2013},
  publisher={Centre for Learning Technology}
}

@article{costley2021effects,
  title={The effects of video lecture viewing strategies on cognitive load},
  author={Costley, Jamie and Fanguy, Mik and Lange, Chris and Baldwin, Matthew},
  journal={Journal of Computing in Higher Education},
  volume={33},
  number={1},
  pages={19--38},
  year={2021},
  publisher={Springer}
}

@article{fabio2015adhd,
  title={ADHD: Auditory and visual stimuli in automatic and controlled processes},
  author={Fabio, Rosa Angela and Castriciano, Claudia and Rondanini, Alessia},
  journal={Journal of Attention Disorders},
  volume={19},
  number={9},
  pages={771--778},
  year={2015},
  publisher={Sage Publications Sage CA: Los Angeles, CA}
}

@article{choe2019student,
  title={Student satisfaction and learning outcomes in asynchronous online lecture videos},
  author={Choe, Ronny C and Scuric, Zorica and Eshkol, Ethan and Cruser, Sean and Arndt, Ava and Cox, Robert and Toma, Shannon P and Shapiro, Casey and Levis-Fitzgerald, Marc and Barnes, Greg and others},
  journal={CBE—Life Sciences Education},
  volume={18},
  number={4},
  pages={ar55},
  year={2019},
  publisher={American Society for Cell Biology}
}

@article{wong2023effectiveness,
  title={Effectiveness of technology-based interventions for school-age children with attention-deficit/hyperactivity disorder: systematic review and meta-analysis of randomized controlled trials},
  author={Wong, Ka Po and Qin, Jing and Xie, Yao Jie and Zhang, Bohan},
  journal={JMIR Mental Health},
  volume={10},
  pages={e51459},
  year={2023},
  publisher={JMIR Publications Toronto, Canada}
}

@inproceedings{lopez2020development,
  title={Development of a home accompaniment system providing homework assistance for children with ADHD},
  author={L{\'o}pez-P{\'e}rez, Laura and Berrezueta-Guzman, Jonnathan and Mart{\'\i}n-Ruiz, Mar{\'\i}a-Luisa},
  booktitle={Conference on Information and Communication Technologies of Ecuador},
  pages={22--35},
  year={2020},
  organization={Springer}
}

@article{meaux2009adhd,
  title={ADHD in the college student: A block in the road},
  author={Meaux, JB and Green, A and Broussard, L},
  journal={Journal of psychiatric and mental health nursing},
  volume={16},
  number={3},
  pages={248--256},
  year={2009},
  publisher={Wiley Online Library}
}

@article{cohen2012importance,
  title={The importance of self-regulation for college student learning},
  author={Cohen, Marisa T},
  journal={College Student Journal},
  volume={46},
  number={4},
  pages={892--903},
  year={2012},
  publisher={Project Innovation Austin LLC}
}

@inproceedings{zhu2026scaffolding,
  title={Scaffolding Metacognition with GenAI: Exploring Design Opportunities to Support Task Management for University Students with ADHD},
  author={Zhu, Zihao and Yu, Junnan and Luo, Yuhan},
  booktitle={Proceedings of the 2026 CHI Conference on Human Factors in Computing Systems},
  pages={1--24},
  year={2026}
}

@article{chen2026not,
  title={" Not Just Me and My To-Do List": Understanding Challenges of Task Management for Adults with ADHD and the Need for AI-Augmented Social Scaffolds},
  author={Chen, Jingruo and Meng, Yibo and Nie, Kexin},
  journal={arXiv preprint arXiv:2603.17258},
  year={2026}
}

@inproceedings{zhang2025understood,
  title={Understood: Real-time communication support for adults with adhd using mixed reality},
  author={Zhang, Shizhen and Li, Shengxin and Li, Quan},
  booktitle={Proceedings of the 38th Annual ACM Symposium on User Interface Software and Technology},
  pages={1--23},
  year={2025}
}

@inproceedings{riaz2024interaction,
  title={Interaction Design Strategies for ADHD Learning Attention—A Review},
  author={Riaz, Hadia and Ullah, Abrar and Zito, Claudio and Soobhany, Ahmed Ryad},
  booktitle={International Conference on Information Technology and Applications},
  pages={321--336},
  year={2024},
  organization={Springer}
}

@article{chen2015effects,
  title={Effects of different video lecture types on sustained attention, emotion, cognitive load, and learning performance},
  author={Chen, Chih-Ming and Wu, Chung-Hsin},
  journal={Computers \& Education},
  volume={80},
  pages={108--121},
  year={2015},
  publisher={Elsevier}
}

@incollection{schmeck1988individual,
  title={Individual differences and learning strategies},
  author={Schmeck, Ronald R},
  booktitle={Learning and study strategies},
  pages={171--191},
  year={1988},
  publisher={Elsevier}
}

@article{alwawi2026beyond,
  title={Beyond learning preferences: exploring the relationship between learning styles and sensory processing among university students},
  author={Alwawi, Dua’a Akram and Madi, Hanan and Abu-Dahab, Sana M N and AlHeresh, Rawan},
  journal={BMC Medical Education},
  year={2026},
  publisher={Springer}
}

@article{vandewaetere2011contribution,
  title={The contribution of learner characteristics in the development of computer-based adaptive learning environments},
  author={Vandewaetere, Mieke and Desmet, Piet and Clarebout, Geraldine},
  journal={Computers in Human Behavior},
  volume={27},
  number={1},
  pages={118--130},
  year={2011},
  publisher={Elsevier}
}

@article{wang2025development,
  title={Development and techniques in learner model in adaptive e-learning system: A systematic review},
  author={Wang, Xiyu and Maeda, Yukiko and Chang, Hua-Hua},
  journal={Computers \& Education},
  volume={225},
  pages={105184},
  year={2025},
  publisher={Elsevier}
}

@article{phobun2010adaptive,
  title={Adaptive intelligent tutoring systems for e-learning systems},
  author={Phobun, Pipatsarun and Vicheanpanya, Jiracha},
  journal={Procedia-Social and Behavioral Sciences},
  volume={2},
  number={2},
  pages={4064--4069},
  year={2010},
  publisher={Elsevier}
}

@article{chen2018recommendation,
  title={Recommendation system for adaptive learning},
  author={Chen, Yunxiao and Li, Xiaoou and Liu, Jingchen and Ying, Zhiliang},
  journal={Applied psychological measurement},
  volume={42},
  number={1},
  pages={24--41},
  year={2018},
  publisher={Sage Publications Sage CA: Los Angeles, CA}
}

@article{sabeima2022towards,
  title={Towards personalized adaptive learning in e-learning recommender systems},
  author={Sabeima, Massra and Lamolle, Myriam and Nanne, Mohamedade Farouk},
  journal={International Journal of Advanced Computer Science and Applications},
  volume={13},
  number={8},
  pages={14--20},
  year={2022},
  publisher={Science and Information (SAI) Organization Limited}
}

@inproceedings{alhosban2024alt,
  title={ALT-D: Enhancing accessibility with an adaptive learning technologies assessment model for students with disabilities},
  author={Alhosban, Amal and Amoush, Rana and Al-Ababneh, Hassan},
  booktitle={2024 IEEE 30th International Conference on Telecommunications (ICT)},
  pages={1--5},
  year={2024},
  organization={IEEE}
}

@article{gevorgyan2024use,
  title={The use of adaptive learning technologies in e-learning for inclusive education: A systematic review},
  author={Gevorgyan, Suren},
  journal={E-Learning Innovations Journal},
  volume={2},
  number={1},
  pages={90--107},
  year={2024}
}

@article{zhang2025influence,
  title={Influence of audiovisual features of short video advertising on consumer engagement behaviors: Evidence from TikTok},
  author={Zhang, Zhipeng and Qiu, Keda and Ye, Yan},
  journal={Journal of Business Research},
  volume={201},
  pages={115662},
  year={2025},
  publisher={Elsevier}
}

@article{mardhatilah2023digital,
  title={Digital consumer engagement: Examining the impact of audio and visual stimuli exposure in social media},
  author={Mardhatilah, Dina and Omar, Azizah and Thurasamy, Ramayah and Juniarti, Rosa Prafitri},
  journal={Information Management and Business Review},
  volume={15},
  number={4},
  pages={94--108},
  year={2023},
  publisher={AMH International}
}

@article{yu2022effects,
  title={Effects of video length on a flipped English classroom},
  author={Yu, Zhonggen and Gao, Mingle},
  journal={Sage Open},
  volume={12},
  number={1},
  pages={21582440211068474},
  year={2022},
  publisher={SAGE Publications Sage CA: Los Angeles, CA}
}

@inproceedings{lee2023lecture,
  title={Lecture presentations multimodal dataset: Towards understanding multimodality in educational videos},
  author={Lee, Dong Won and Ahuja, Chaitanya and Liang, Paul Pu and Natu, Sanika and Morency, Louis-Philippe},
  booktitle={Proceedings of the IEEE/CVF International Conference on Computer Vision},
  pages={20087--20098},
  year={2023}
}

@inproceedings{manasrah2021short,
  title={Short videos, or long videos? A study on the ideal video length in online learning},
  author={Manasrah, Ahmad and Masoud, Mohammad and Jaradat, Yousef},
  booktitle={2021 international conference on information technology (ICIT)},
  pages={366--370},
  year={2021},
  organization={IEEE}
}

@article{galotti2019students,
  title={Students choosing courses: Real-life academic decision making},
  author={Galotti, Kathleen M and Umscheid, Valerie A},
  journal={The American Journal of Psychology},
  volume={132},
  number={2},
  pages={149--159},
  year={2019},
  publisher={University of Illinois Press}
}

@inproceedings{kumar2008improving,
  title={Improving the accuracy of gaze input for interaction},
  author={Kumar, Manu and Klingner, Jeff and Puranik, Rohan and Winograd, Terry and Paepcke, Andreas},
  booktitle={Proceedings of the 2008 symposium on Eye tracking research \& applications},
  pages={65--68},
  year={2008}
}

@misc{pyscenedetect,
  author       = {Brandon Castellano},
  title        = {PySceneDetect},
  howpublished = {\url{https://www.scenedetect.com/}},
  year         = {2025}
}

@inproceedings{wang2025characterizing,
  title={Characterizing Visual Intents for People with Low Vision through Eye Tracking},
  author={Wang, Ru and Chen, Ruijia and Cai, Anqiao Erica and Li, Zhiyuan and Mondal, Sanbrita and Zhao, Yuhang},
  booktitle={Proceedings of the 27th International ACM SIGACCESS Conference on Computers and Accessibility},
  pages={1--18},
  year={2025}
}

@article{zar2005spearman,
  title={Spearman rank correlation},
  author={Zar, Jerrold H},
  journal={Encyclopedia of biostatistics},
  volume={7},
  year={2005},
  publisher={Wiley Online Library}
}

@article{negi2020fixation,
  title={Fixation duration and the learning process: An eye tracking study with subtitled videos},
  author={Negi, Shivsevak and Mitra, Ritayan},
  journal={Journal of Eye Movement Research},
  volume={13},
  number={6},
  pages={40},
  year={2020},
  publisher={Bern Open Publishing}
}

@article{borys2017eye,
  title={Eye-tracking metrics in perception and visual attention research},
  author={Borys, Magdalena and Plechawska-W{\'o}jcik, Ma{\l}gorzata},
  journal={EJMT},
  volume={3},
  number={16},
  pages={11--23},
  year={2017}
}

@article{wang2021multi,
  title={Multi-sensor eye-tracking systems and tools for capturing student attention and understanding engagement in learning: A review},
  author={Wang, Yuehua and Lu, Shulan and Harter, Derek},
  journal={IEEE Sensors Journal},
  volume={21},
  number={20},
  pages={22402--22413},
  year={2021},
  publisher={IEEE}
}

@article{reichle2010eye,
  title={Eye movements during mindless reading},
  author={Reichle, Erik D and Reineberg, Andrew E and Schooler, Jonathan W},
  journal={Psychological science},
  volume={21},
  number={9},
  pages={1300--1310},
  year={2010},
  publisher={Sage Publications Sage CA: Los Angeles, CA}
}

@article{moiroud2025gaze,
  title={Gaze Dispersion During a Sustained-Fixation Task as a Proxy of Visual Attention in Children with ADHD},
  author={Moiroud, Lionel and Moscoso, Ana and Acquaviva, Eric and Michel, Alexandre and Delorme, Richard and Bucci, Maria Pia},
  journal={Vision},
  volume={9},
  number={3},
  pages={76},
  year={2025},
  publisher={MDPI}
}

@article{krasich2020eyes,
  title={Where the eyes wander: The relationship between mind wandering and fixation allocation to visually salient and semantically informative static scene content},
  author={Krasich, Kristina and Huffman, Greg and Faber, Myrthe and Brockmole, James R},
  journal={Journal of vision},
  volume={20},
  number={9},
  pages={10--10},
  year={2020},
  publisher={The Association for Research in Vision and Ophthalmology}
}

@article{geissler2014hyperactivity,
  title={Hyperactivity and sensation seeking as autoregulatory attempts to stabilize brain arousal in ADHD and mania?},
  author={Geissler, Julia and Romanos, Marcel and Hegerl, Ulrich and Hensch, Tilman},
  journal={ADHD Attention Deficit and Hyperactivity Disorders},
  volume={6},
  number={3},
  pages={159--173},
  year={2014},
  publisher={Springer}
}

@article{lackmann2021influence,
  title={The influence of video format on engagement and performance in online learning},
  author={Lackmann, Sergej and L{\'e}ger, Pierre-Majorique and Charland, Patrick and Aub{\'e}, Caroline and Talbot, Jean},
  journal={Brain Sciences},
  volume={11},
  number={2},
  pages={128},
  year={2021},
  publisher={MDPI}
}

@article{kasneci2023chatgpt,
  title={ChatGPT for good? On opportunities and challenges of large language models for education},
  author={Kasneci, Enkelejda and Se{\ss}ler, Kathrin and K{\"u}chemann, Stefan and Bannert, Maria and Dementieva, Daryna and Fischer, Frank and Gasser, Urs and Groh, Georg and G{\"u}nnemann, Stephan and H{\"u}llermeier, Eyke and others},
  journal={Learning and individual differences},
  volume={103},
  pages={102274},
  year={2023},
  publisher={Elsevier}
}

@inproceedings{reihanian2024review,
  title={A Review of Generative AI in Computer Science Education: Challenges and Opportunities in Accuracy, Authenticity, and Assessment},
  author={Reihanian, Iman and Hou, Yunfei and Chen, Yu and Zheng, Yifei},
  booktitle={International Conference on Computational Science and Computational Intelligence},
  pages={144--158},
  year={2024},
  organization={Springer}
}

@article{zhang2024vision,
  title={Vision-language models for vision tasks: A survey},
  author={Zhang, Jingyi and Huang, Jiaxing and Jin, Sheng and Lu, Shijian},
  journal={IEEE transactions on pattern analysis and machine intelligence},
  volume={46},
  number={8},
  pages={5625--5644},
  year={2024},
  publisher={IEEE}
}

@inproceedings{baclawski2018observer,
  title={The observer effect},
  author={Baclawski, Kenneth},
  booktitle={2018 ieee conference on cognitive and computational aspects of situation management (cogsima)},
  pages={83--89},
  year={2018},
  organization={IEEE}
}

@article{weinstein2018mind,
  title={Mind-wandering, how do I measure thee with probes? Let me count the ways},
  author={Weinstein, Yana},
  journal={Behavior research methods},
  volume={50},
  number={2},
  pages={642--661},
  year={2018},
  publisher={Springer}
}

@inproceedings{srivastava2019continuous,
  title={Continuous evaluation of video lectures from real-time difficulty self-report},
  author={Srivastava, Namrata and Velloso, Eduardo and Lodge, Jason M and Erfani, Sarah and Bailey, James},
  booktitle={Proceedings of the 2019 CHI conference on Human factors in computing systems},
  pages={1--12},
  year={2019}
}

@article{peel2019fundamentals,
  title={The fundamentals for self-regulated learning: A framework to guide analysis and reflection},
  author={Peel, Karen},
  journal={Educational Practice and theory},
  volume={41},
  number={1},
  pages={23--49},
  year={2019},
  publisher={James Nicholas Publishers}
}

@article{schunk2005self,
  title={Self-regulated learning: The educational legacy of Paul R. Pintrich},
  author={Schunk, Dale H},
  journal={Educational psychologist},
  volume={40},
  number={2},
  pages={85--94},
  year={2005},
  publisher={Taylor \& Francis}
}

@book{ausubel2012acquisition,
  title={The acquisition and retention of knowledge: A cognitive view},
  author={Ausubel, David Paul},
  year={2012},
  publisher={Springer Science \& Business Media}
}

@article{abbasi2014measuring,
  title={Measuring effectiveness of learning chatbot systems on student’s learning outcome and memory retention},
  author={Abbasi, Suhni and Kazi, Hameedullah},
  journal={Asian Journal of Applied Science and Engineering},
  volume={3},
  number={2},
  pages={251--260},
  year={2014}
}

@article{ashby2004monitoring,
  title={Monitoring student retention in the Open University: Definition, measurement, interpretation and action},
  author={Ashby*, Alison},
  journal={Open Learning: The Journal of Open, Distance and e-Learning},
  volume={19},
  number={1},
  pages={65--77},
  year={2004},
  publisher={Taylor \& Francis}
}


\end{document}